\documentclass{article}

\usepackage{arxiv}

\usepackage[utf8]{inputenc} 
\usepackage[T1]{fontenc}    
\usepackage{hyperref}       
\usepackage{url}            
\usepackage{booktabs}       
\usepackage{amsfonts}       
\usepackage{nicefrac}       
\usepackage{microtype}      
\usepackage{lipsum}
\usepackage{subcaption}
\usepackage{graphicx}
\usepackage{siunitx}
\graphicspath{ {./figures/} }
\usepackage[colorinlistoftodos]{todonotes}

\usepackage{epstopdf}
\usepackage{authblk}
\DeclareGraphicsExtensions{.png,.pdf}

\title{Biases in human mobility data impact epidemic modeling}

\author[1, 2]{Frank Schlosser}
\author[3, 4]{Vedran Sekara}
\author[1, 2]{Dirk Brockmann}
\author[4]{Manuel Garcia-Herranz}

\affil[1]{Robert Koch-Institute, Nordufer 20, D-13353 Berlin, Germany}
\affil[2]{Institute for Theoretical Biology, Humboldt-University of Berlin, Philippstr. 13, D-10115 Berlin, Germany}
\affil[3]{IT University of Copenhagen, Rued Langgaards Vej 7, 2300 Copenhagen, Denmark}
\affil[4]{UNICEF, 3 United Nations Plaza, New York, NY 10017, USA}

\begin{document}
\maketitle

\begin{abstract}
Large-scale human mobility data is a key resource in data-driven policy making and across many scientific fields.
Most recently, mobility data was extensively used during the COVID-19 pandemic to study the effects of governmental policies and to inform epidemic models.
Large-scale mobility is often measured using digital tools such as mobile phones. However, it remains an open question how truthfully these digital proxies represent the actual travel behavior of the general population.
Here, we examine mobility datasets from multiple countries and identify two fundamentally different types of bias caused by unequal access to, and unequal usage of mobile phones.
We introduce the concept of \emph{data generation} bias, a previously overlooked type of bias, which is present when the amount of data that an individual produces influences their representation in the dataset.
We find evidence for data generation bias in all examined datasets in that high-wealth individuals are overrepresented, with the richest 20\% contributing over 50\% of all recorded trips, substantially skewing the datasets.
This inequality is consequential, as we find mobility patterns of different wealth groups to be structurally different, where the mobility networks of high-wealth users are denser and contain more long-range connections.
To mitigate the skew, we present a framework to debias data and show how simple techniques can be used to increase representativeness.
Using our approach we show how biases can severely impact outcomes of dynamic processes such as epidemic simulations, where biased data incorrectly estimates the severity and speed of disease transmission.
Overall, we show that a failure to account for biases can have detrimental effects on the results of studies and urge researchers and practitioners to account for data-fairness in all future studies of human mobility.

\end{abstract}


\section*{Introduction}
A large range of applications, from urban planning~\cite{wang2012understanding}, dynamic population mapping~\cite{deville2014dynamic}, estimation of migrations ~\cite{palotti2020monitoring}, to epidemic modeling~\cite{wesolowski2014commentary} rely on large-scale human mobility datasets. 
Mobility data has even been proposed as a cost-effective way of estimating population level statistics in low- and middle-income countries~\cite{blumenstock2015predicting,pokhriyal2017combining}.
More recently, human mobility data has gained massive attention in the wake of the COVID-19 pandemic~\cite{oliver2020mobile}, where it has been used to monitor the impact of mobility restrictions world-wide~\cite{kraemer2020effect, Flaxman2020, galeazzi2020human}, to understand the complex effects of governmental policies~\cite{bonaccorsi2020economic, schlosser2020covid, chang2020mobility, Dehning2020}, and as a key ingredient for epidemiological modelling~\cite{chinazzi2020effect, Gatto2020, arenas2020mathematical}.
For all these applications it is crucial to have unbiased estimates of human travels that accurately represent the behavior of the underlying populations.

In our increasingly digital world, large scale mobility data is often passively collected as a by-product of digital technologies used for billing, service, or marketing purposes.
This includes call detail records (CDRs) collected through regular mobile phone usage~\cite{gonzalez2008understanding}, GPS traces collected via smartphone apps~\cite{stopczynski2014measuring}, check-ins from online social media services~\cite{jurdak2015understanding}, and smart travel card data~\cite{batty2013big} to mention a few.
However, it is an open question how well these proxy datasets capture the true movements of people.
It is widely recognized that access to, and usage of, big-data technologies is heterogeneous across populations~\cite{james2007mobile,lazer2014parable,sapiezynski2020fallibility}.
Differences in technology usage can lead to disparities in how, or whether, individuals are captured in digitally collected datasets~\cite{sekara2019mobile}.
For instance, it has been established that in certain countries mobile phone ownership is biased towards predominantly wealthier, better educated, and for the most part male populations~\cite{blumenstock2010mobile,wesolowski2012heterogeneous}, meaning that mobility datasets captured in this context will mainly contain the travels of these demographics.
Unequal access to digital technologies is especially troubling as it is often deeply intertwined with socioeconomic, geographic, ethnic, gender, ability, and class disparities~\cite{wesolowski2013impact}.

The question of representation, i.e. whose behavior do digitally collected datasets represent, is a pressing matter as uncorrected biases in mobility data can lead to incorrect conclusions with potentially harmful adverse impacts.
For example, in the context of the current COVID-19 pandemic, government policies have in many countries been based on epidemiological models which use human mobility as a key ingredient.
Mis- and underrepresention of specific population demographics, however, could lead to discriminatory policies which disadvantage populations that are not properly captured in the data.
This is especially problematic as researchers often use pre-computed datasets shared by third party data providers or other entities, and thus have only limited insight into the methodology of data collection, leaving potential sources of bias unnoticed.
Despite the potential harmful impacts of biases, few studies have systematically investigated the presence of biases in human mobility datasets and what effects they can have on dynamic processes simulated on these datasets, such as epidemic models. 

In this study we examine mobility datasets estimated from call detail records (CDRs) from 3 countries (Sierra Leone, the Democratic Republic of the Congo (DRC), and Iraq) for different forms of bias and we report a new type of bias which has not previously been reported in human mobility data.
In addition to the previously well-documented \textit{technology access} bias (not everybody has access to digital technologies such as mobile phones), we find evidence of a fundamentally different type of bias --- \textit{data generation} bias --- where different demographics produce, or generate, unequal amounts of data.

Access to technology bias concerns the question of how access to the recording technology is distributed across different demographics.
This type of bias is related to ownership of mobile devices and is well documented in the literature, where gender, age, disability, education, ethnicity, and wealth characteristics have been identified as factors that influence ownership~\cite{blumenstock2010mobile,wesolowski2012heterogeneous,wesolowski2013impact,aranda2019understanding}.
As such, there exist methodologies to both quantify and partially reduce this type of bias in mobility data~\cite{wesolowski2013impact, tizzoni2014use, coston2020leveraging, pestre2020abcde}.

Data generation bias is fundamentally different, it deals not with \textit{if}, but \emph{how} a user is captured in a dataset.
Even with homogeneous societal access to the recording technology (in this case mobile phones), equal representation is not guaranteed if technology usage depends on the socioeconomic properties of individuals.
If an activity has an associated cost, for instance the cost of making a phone call, poorer individuals might limit their activity to save money, in turn lowering their representation in the collected data.
For human mobility derived from mobile phones, data generation bias will manifest in individuals with low mobile phone activity (few calls and text messages) having less trips captured in the data (see Fig.~\ref{fig:bias_in_data}a).
In fact, previous studies have demonstrated that low mobile phone activity (low number of calls and texts) is reflected in the resolution with which individual mobility patterns can be reconstructed~\cite{ranjan2012call, zhao2016understanding}, even if an individual has undertaken the same amount of trips as a person with high phone activity.

We demonstrate the presence of both technology access and data generation bias in the examined datasets and present a framework to correct for, or \emph{debias}, mobility datasets for these biases in order to improve the representativeness in terms of socioeconomic characteristics.
Using epidemic simulations as an example, we compare the outcome of epidemic models running on biased and debiased data and demonstrate that biases can severely impact the result of dynamic processes.
As such, if biases are not taken into account they can greatly influence insights derived from epidemic simulations.

\section*{Results}

We analyze CDR estimated mobility data from Sierra Leone, Democratic Republic of the Congo (DRC), and Iraq (see Supplementary Materials for a full description of the datasets).
Data is generated in the form of mobile phone activity (texts or calls), which is registered at the closest cell tower.
Trips are then recorded as movements between cell towers.
The data processing differs slightly among the datasets used here: For the Sierra Leone and DRC datasets, trips take place between subsequent locations of data activity; For Iraq, a trip is counted from the users home location to the location of data activity (see SI for details).
Trips are estimated on an individual level, and, to preserve privacy, aggregated by the mobile networks operators (MNOs) spatially and temporally to create the mobility networks $F$ for each country.
The flow $F_{ij}$ quantifies the total number of trips from region $i$ to $j$, recorded within the time frame of the dataset, including flows $F_{ii}$ that start and end in the same region.
Here, the regions are the $m$ districts in each country (or their corresponding administrative level 3 division, see SI for details).

To understand the impact of socioeconomic status of users on data generation and on mobility patterns, we study subsets of the mobility networks $F$ which are distinguished by wealth.
Before the datasets are aggregated the MNOs split up the users in each dataset into 5 equally sized groups (quintiles $q \in \left\{\mathrm{Q1},\ldots,\mathrm{Q5} \right\}$) according to their airtime expenditure, with Q1 having the lowest and Q5 the highest expenditure (see Fig.~\ref{fig:bias_in_data}b).
Airtime expenditure has been previously established to to correlate well with underlying wealth~\cite{wesolowski2013impact} and food security~\cite{decuyper2014estimating}.
We thus classify the quintile Q1 (Q5) of users with the lowest (highest) airtime expenditure to have the lowest (highest) socioeconomic status, and refer to them as poorer (richer) users (see SI section 1.1 for more details).
The MNOs provided us with separate mobility networks $F^{(q)}$, each containing only trips of users in quintile $q$.
The original, aggregate network $F$ is the union of these quintile networks $F^{(q)}$, where the flows are added up to the total flow $F_{ij}=\sum_q F_{ij}^{(q)}$.

\subsection*{Imbalances in data generation across wealth quintiles}

\begin{figure}[htb]
\centering
\includegraphics[width=0.9\linewidth]{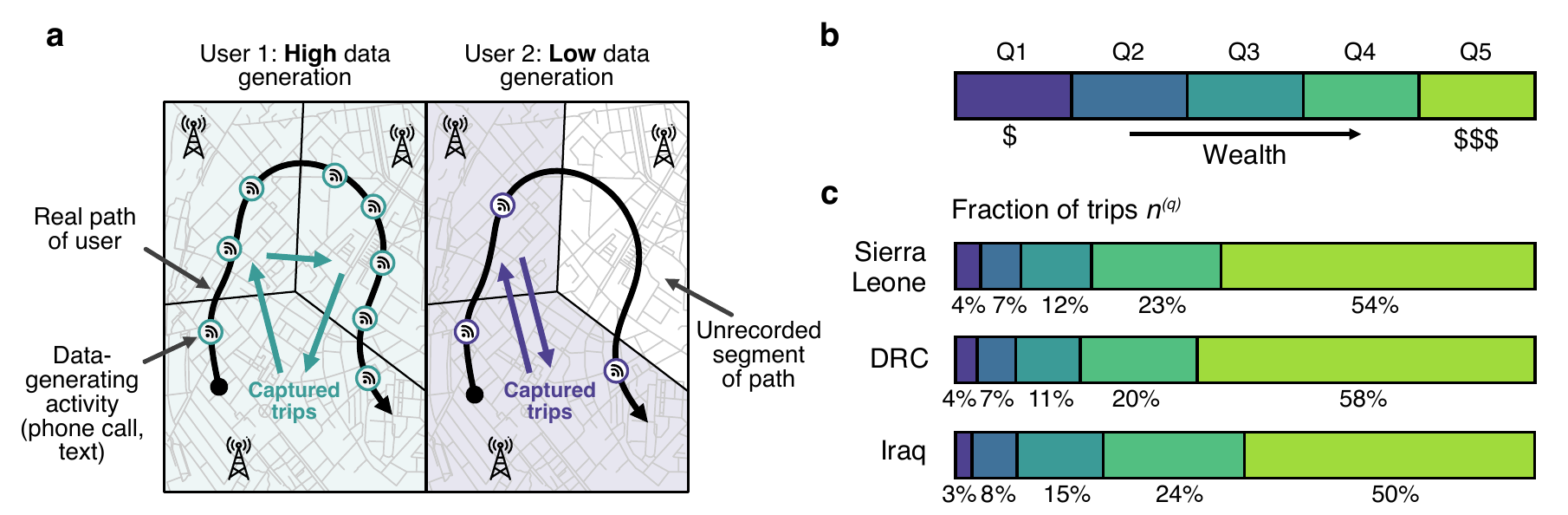}
\caption{\textbf{Illustration of data generation bias and evidence from data.} \textbf{a.} Users 1 and 2 have an identical real path (black arrow), but user 1 has a higher data generation rate leading to more recorded phone activities (white circles). As a result, for user 1 more trips between cell tower regions are captured in the CDR mobility dataset (colored arrows). \textbf{b.} Mobility networks are divided into 5 sub-networks $F^{(q)}$, each representing a quintile or 20\% of the total user base. The users are sorted by their socioeconomic status as measured by airtime expenditure, so that the quintiles $q$ range from the poorest group of users Q1 to the wealthiest group Q5. \textbf{c.} Empirical evidence from three countries shows large inequalities in how users are represented in mobility data. Shown are the fraction of trips $n^{(q)}$ in each quintile among all trips. 
Richer groups contribute more trips and are thus over-represented in all countries. In effect, the top quintile of users (Q5) accounts for roughly half of all trips, while to poorest (Q1) accounts for less than 5\%.}
\label{fig:bias_in_data}
\end{figure}

We find that the wealth of users has an impact on the amount of mobility data they generate, as measured by the number of trips captured in the dataset.
We calculate the fraction of trips $n^{(q)}$ generated by the users in each quintile $q$, as $n^{(q)} = N^{(q)}/N,$ where $N^{(q)}$ is the total number of trips in each quintile network, $N^{(q)}=\sum_{ij} F_{ij}^{(q)}$, and $N$ the number of trips in the full network, $N=\sum_{q} N^{(q)}$.
Fig.~\ref{fig:bias_in_data}c shows there are large inequalities in data representativeness: high-wealth users are overrepresented, with the wealthiest 20\% of users (Q5) contributing approximately 50\% of all recorded trips, while the poorest 20\% (Q1) produce less than 5\% of all trips.
Taken together, the bottom 80\% of users produce approximately the same amount of data as the wealthiest 20\%, a finding which is consistent across countries.
(Similar distributions have been identified across a multitude of systems and are often called Pareto distributions~\cite{newman2005power}.)
This is an important observation as the travel behavior of wealthy individuals dominates the captured mobility data.

\subsection*{The effects of bias on the structure of mobility networks}

\begin{figure}[htb]
\centering
\includegraphics[width=0.95\linewidth]{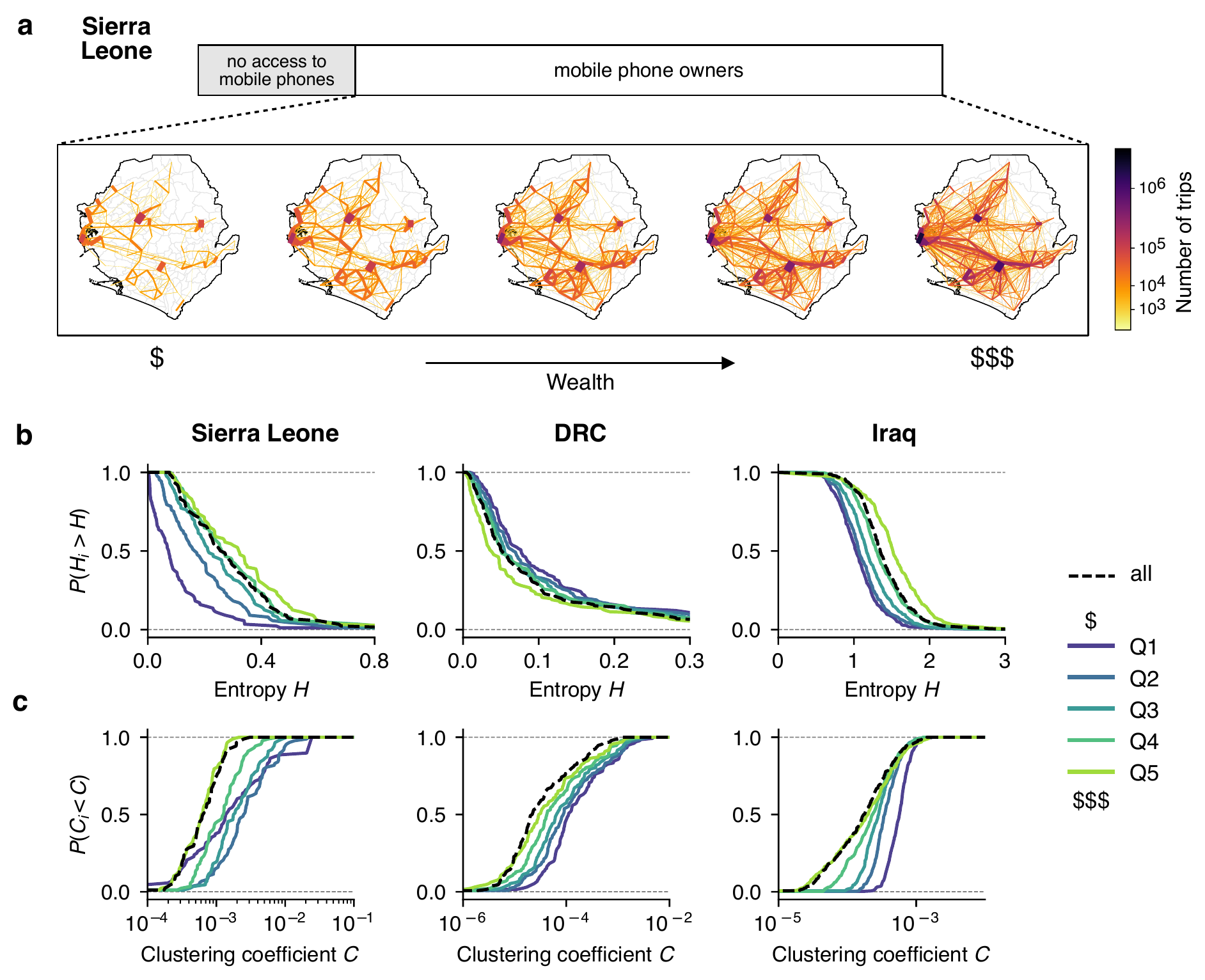}
\caption{\textbf{The effects of data generation bias on the structure of mobility networks.} \textbf{a.} The mobility networks for Sierra Leone for each socioeconomic quintile. The network for wealthier quintiles (Q5) is denser and contains more unique, more long-distance, and more high-flow connections. Edge color and width are scaled according to the number of trips. At the time of data collection there were in Sierra Leone 78.9 mobile phone subscribers per 100 people~\cite{international2020itu}. \textbf{b.} The cumulative distribution of the Shannon entropy $H_i^{(q)}$ of outgoing trips from districts $i$ for different quintiles. For Sierra Leone and Iraq, richer quintiles have a higher entropy indicating more diverse mobility connections. The relationship is weaker and reversed for DRC. \textbf{c.} The cumulative distribution of the weighted clustering coefficient $C_i$ of districts $i$ in each quintile network. The clustering coefficients are generally higher for low-wealth quintiles (Q1), meaning that they are more locally dense with fewer long-distance connections.}
\label{fig:networks_statistics}
\end{figure}



Imbalances in data generation cause a profound over-representation of wealthy individuals, which matters, because as we show in the following, mobility patterns of the quintiles differ.
As a consequence, the aggregate travel network is skewed towards the mobility patterns of rich people, and as we demonstrate in the last section, this distorts the outcomes of dynamic simulations, such as the conclusions we can draw from epidemic models.

In Fig.~\ref{fig:networks_statistics}a we compare the mobility networks $F^{(q)}$ of low- and high-wealth users and find distinct differences between them--- large enough to be visually observed.
The mobility networks of wealthier users contain more trips overall (comparing Q1 and Q5 for Sierra Leone, we find $N^{\mathrm{Q1}}=47$ million and $N^{\mathrm{Q5}}=589$ million), have more unique links $L$  ($L^\mathrm{Q1}= 452$ and $L^\mathrm{Q5}= 1,610$) but the same number of regions $M=88$, and thus have a higher density $\rho=L/M$  ($\rho^\mathrm{Q1}= 5.9\%$ and $\rho^\mathrm{Q5}= 21.0\%$).
In addition, wealthier quintiles have relatively more connections over long-distances, and have a higher proportion of edges with a high number of trips (see results in SI section 2).

In addition, we find significant differences between the quintiles in more advanced topological measures, namely the district-wise Shannon entropy $H_i^{(q)}$ (Fig.~\ref{fig:networks_statistics}b) and weighted clustering coefficient $C_i^{(q)}$ (Fig.~\ref{fig:networks_statistics}c, see definitions in methods).
Shannon entropy is a measure of diversity of the mobility connections $F^{(q)}_{ij}$ starting in district $i$.
It has previously been shown to be connected to the socioeconomic status of a region: Locations with a higher socioeconomic status in general have a more diverse set of connections, corresponding to a higher entropy \cite{Eagle2010, wesolowski2012heterogeneous}.
The relationship is not universal though ~\cite{xu2018human} and can be reversed depending on the spatial organization of cities; for example, in certain places low-income households can be located in the outskirts of cities, and in other places they can be located downtown~\cite{barbosa2021uncovering}.
In our data, we find significant differences in the distributions of entropies between quintiles, see Fig.~\ref{fig:networks_statistics}b.
In Sierra Leone and Iraq, richer quintiles have a higher entropy, indicating a more diverse connections, while the relationship is reversed for DRC.
In all countries, the distributions for Q1 and Q5 are significantly different ($p<0.001$, two-sample Kolmogorov-Smirnov test, see detailed statistics in SI).


Similarly, we find that the distributions of the weighted clustering coefficient differ between quintiles, see Fig.~\ref{fig:networks_statistics}c.
The clustering coefficient measures the average flow between triplets of neighboring districts, where a large value indicates that two neighbors of a district are likely to have large flows between them, too.
We find that poorer quintiles generally have higher clustering coefficients, indicating that these mobility networks are locally more dense, whereas higher wealth network are less locally dense, coinciding with them having a higher proportion of long distance trips as stated above.
Again, differences between Q1 and Q5 are significant in a KS test ($p<0.001$, see SI).
Importantly, the differences in network structure across poor and wealthy groups persist even when we account for imbalances in the total number of trips between the quintiles.
We show this by calculating the metrics for resampled networks $F'^{(q)}$, which all contain the same total number of trips, and find that the observed metrics are qualitatively unaffected (SI, Fig SI2).

In addition to the observed data generation bias, we also find evidence of technology access bias in the mobility networks, meaning that only a fraction of the population is present in the dataset at all.
This is frequently observed in mobile phone data as no single MNO has a monopoly, and commercial factors play a large role on which regions and populations are covered.
We quantify this bias using the technology coverage rate $c_i=U_i/P_i$, which is the number of users $U_i$ of the MNO among the population $P_i$ in district $i$.
Similar to previous studies \cite{tizzoni2014use, coston2020leveraging}, we find that $c_i$ is on average smaller than 1, that it is varies considerably across districts, and that some districts are not present in the network at all, i.e. with $c_i=0$ and no in- or outgoing flows (see results in SI sec.4).
Taken together, data generation and technology access biases hamper our ability to truthfully reconstruct the true mobility network, and choosing not to correct for these biases will create datasets that over-represent the behavior of wealthier groups.

\subsection*{Debiasing mobility data}
\begin{figure}[tb]
\centering
\includegraphics[width=0.9\linewidth]{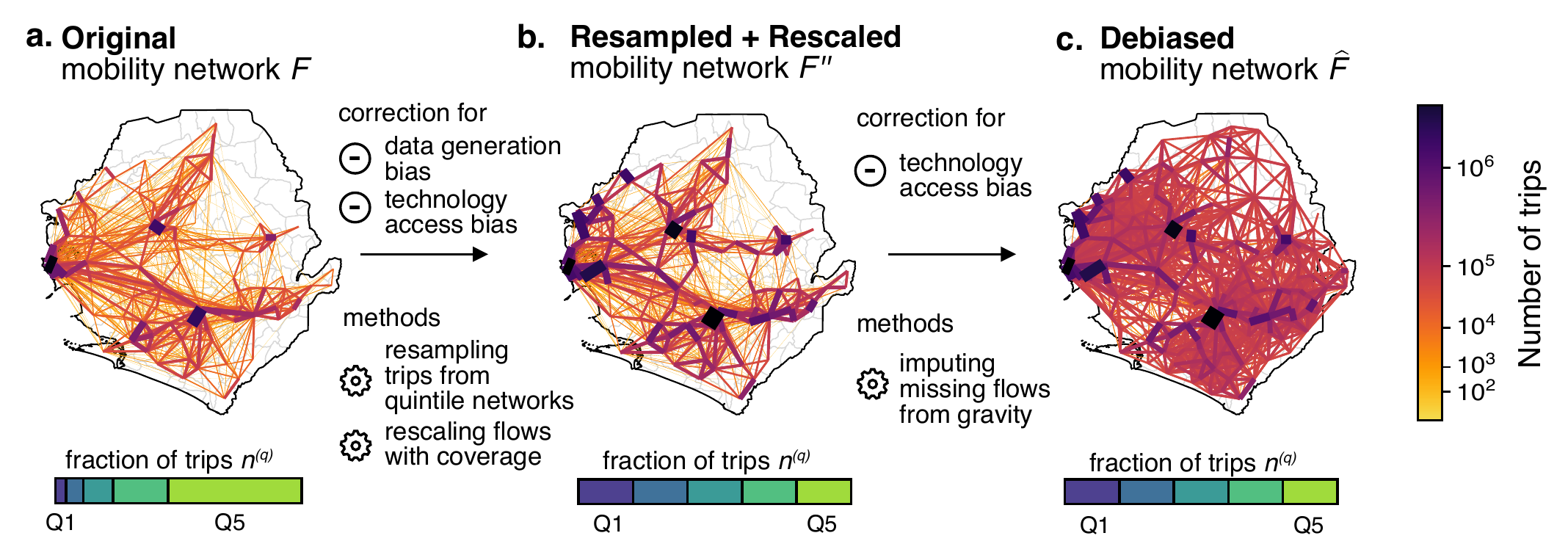}
\caption{\textbf{Illustration of the debiasing process.} \textbf{a.} The original mobility network $F$ exhibits biases regarding technology access (apparent in that some regions entirely lack data) and unequal data generation (wealth quintiles contribute an unequal fraction of trips, illustrated by bars below maps). \textbf{b.} We account for data generation bias by resampling trips from the original mobility network, where we over- and under-sample previously under- and over-represented quintiles, such that a resampled network has an equal fraction of trips from all wealth groups. In addition, we partially correct for technology access bias by rescaling existing flows, accounting for the fact that MNOs have only a fraction of the population as customers, resulting in the resampled and rescaled network $F''$ \textbf{c.} Finally, for regions where no travel data is captured at all ($c_i=0$), we impute flows from a gravity model, resulting in a realization of the debiased mobility network $\widehat{F}$.}
\label{fig:debiasing_illustration}
\end{figure}

We formulate a methodology to construct debiased mobility networks $\widehat{F}$ that accounts for the data generation and technology access bias present in the original network $F$, see Fig.~\ref{fig:debiasing_illustration}.
The debiased network is our estimation of what ``unbiased mobility data'' would look like in the optimal situation, where all users in the population had equal access to and equal means of data production.
Our methodology is based on a general mathematical debiasing framework which can be adapted to other specific contexts (the full framework is described in SI sec.~3).

The quintile networks $F^{\prime (q)}$ contain very different amounts of trips $N^{(q)}$ (see Fig.~\ref{fig:bias_in_data}), leading to an unequal representation in the aggregate network $F$.
To mitigate for the skew in data generation we construct resampled quintile networks $F'^{(q)}$ by sampling trips from the original quintile networks so that each graph contains the same amount of trips $N^*$, i.e. we over- or undersample the previously under- and overrepresented networks (see Materials and Methods).
The resampled networks $F'^{(q)}$ are then combined to form a resampled aggregate network $F'=\sum_q F'^{(q)}$, where each quintile of users contributes the same amount of trips.

In addition to the resampling, we also use established techniques \cite{tizzoni2014use} to account for technology access bias, i.e. for users and regions which are not represented in the dataset at all.
First, in districts where mobility data is present, but where only a fraction $c_i<1$ of the total population are users of the MNO, we rescale all flows $F'_{ij}$ starting in the district $i$ with the coverage $c_i$ to estimate the mobility of the total population,
$F^{\prime\prime}_{ij}=F^{\prime}_{ij}/c_i$, yielding the resampled and rescaled network $F''$.
Second, for districts where there is no coverage at all ($c_i=0$), and where there consequently is no recorded mobility data, $F'_{ij}=F'_{ji}=0$, we estimate the missing flows starting and ending in this district from a gravity-like human mobility model (see details in Materials and Methods).

The end result is a realization of the debiased network $\widehat{F}$, which corrects for data generation and technology access biases.
The average debiased network differs from the original network $F$ in many ways, most notably in that it covers the full area of the country and overall includes more trips (Fig.~\ref{fig:debiasing_illustration}).
Further, trips are more evenly distributed among the edges and nodes in the debiased networks, indicating a more equal representation of districts (see results in SI section 4).

\subsection*{How biased mobility data effects epidemiological predictions}

Unaddressed biases in mobility datasets are dangerous because they can affect the outcome of dynamic processes based on mobility, which we demonstrate using the example of epidemic spreading.
We simulate epidemics on the original mobility network $F$ as well as on the debiased network $\widehat{F}$ and compare the differences.
We use an SIR model to simulate a contagion process in metapopulations that are connected by a commuter-type mobility~\cite{schlosser2020covid,tizzoni2014use}.
For each simulation, we start the epidemic with a seed of infecteds in a randomly chosen districts present in the original data (for details see Materials and Methods).

\begin{figure}[!htb]
\centering
\includegraphics[width=0.85\linewidth]{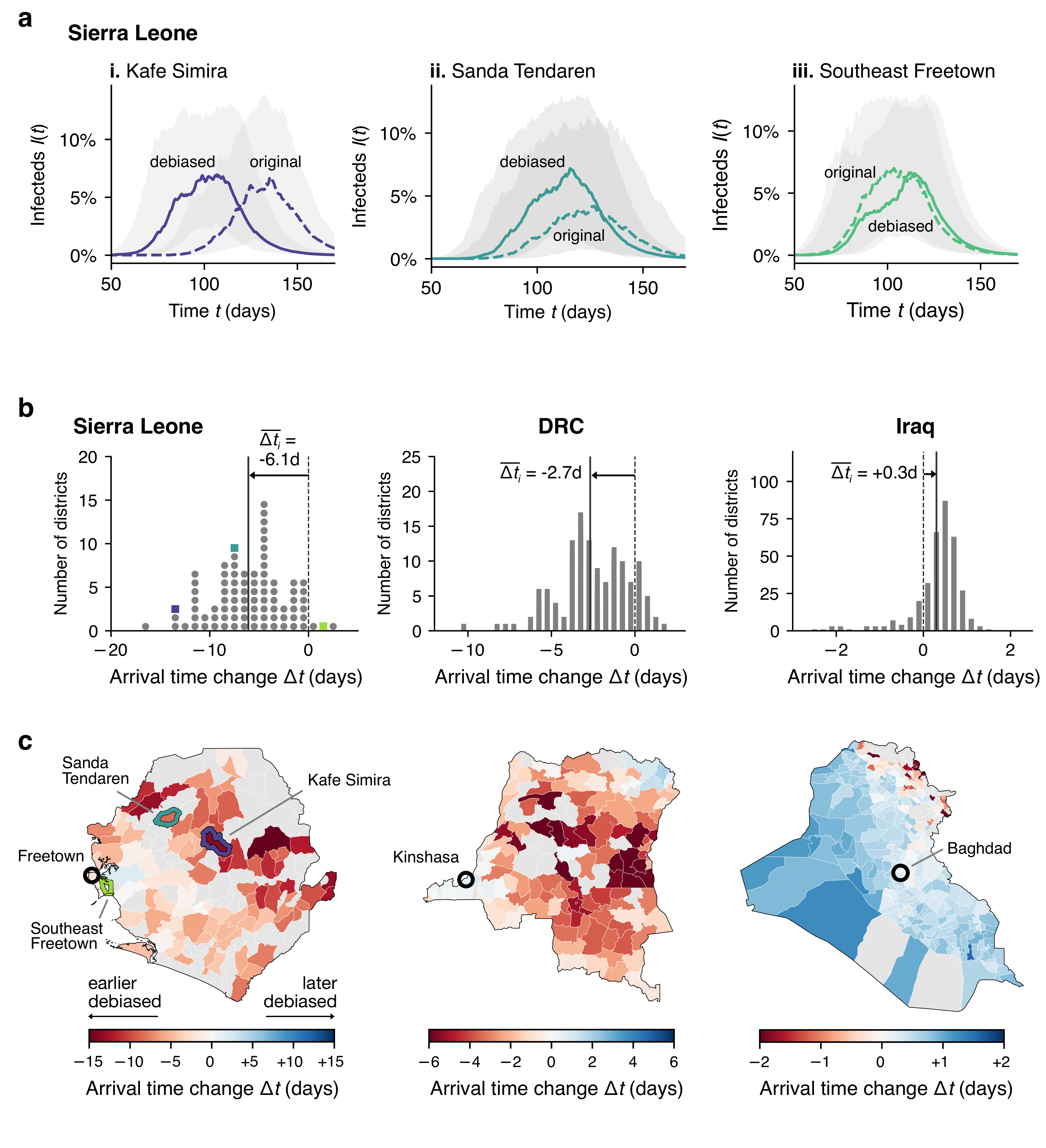}
\caption{\textbf{Effects of biases on epidemic spreading}. We simulate and compare epidemics on the original mobility network and realizations of the debiased network. Epidemics are started in random districts present in both datasets and results are averaged over multiple realizations (see Methods and Materials). \textbf{a.} The fraction of infected $I(t)$ in three sample districts of Sierra Leone over time. Depending on the district, the epidemic curve can be shifted to earlier or later times, have a higher or lower peak, or an entirely different shape, which shows the diverse impact of biases on a sub-national level. Lines show the median of $I(t)$, and shaded areas encompass the most central $50\%$ of curves (see Materials and Methods). \textbf{b.} We find a substantial variation in the impact of biases on districts, as shown by the wide distribution of the arrival time changes $\Delta t_i= \langle t_i\rangle - \langle \widehat{t}_i \rangle$ in districts $i$, which is the change in arrival time when switching to the debiased dataset.
\textbf{c.} Changes in arrival times are spatially heterogeneous, with the epidemic often arriving earlier in remote, low-populated areas. Black circles depict national capitals.}
\label{fig:epi_results}
\end{figure}

On a sub-national level, we find stark differences between the spreading pattern of the epidemic on the debiased mobility networks compared to the original network, see Fig.~\ref{fig:epi_results}.
Depending on the district, the disease can arrive both substantially earlier or later than predicted, and the epidemic curve can have an entirely different shape (Fig.~\ref{fig:epi_results}a).
To quantify the spatially heterogeneous impact of debiasing, we calculate the arrival time change $\Delta t_i= \langle t_i\rangle - \langle \widehat{t}_i \rangle$ for each district as the difference in the average arrival time $\langle t_i \rangle$ in the original mobility network (averaged across simulations) compared to the average arrival time $\langle \widehat{t}_i \rangle$ in the debiased mobility networks.
We find a strong variation in the arrival time change $\Delta t_i$ across districts (Fig.~\ref{fig:epi_results}b).
In Sierra Leone, the epidemic arrives on average 6 days earlier across all districts ($-11.3 \pm 1.8\%$ change), but for some districts the virus can arrive up to 16 days earlier than predicted based on the original data ($-26.8 \pm 1.5\%$ change).
Results for DRC and Iraq show similar discrepancies for arrival time changes, although less extreme in magnitude, on average 2.7 days before for DRC, and 0.3 days later for Iraq.
A spatial investigation of the arrival time changes reveals that the districts where the epidemic arrives earlier are often remote and low populated areas (Fig.~\ref{fig:epi_results}c).
On a national level, the epidemic curves show a smaller change when debiasing the mobility data (see additional results in SI), which highlights that the heterogeneous impact of biases on a regional level can be missed when focusing on national averages.


\newpage
\section*{Discussion}
\label{ch:discussion}

The mobile phone datasets used in many fields as a proxy for large-scale human mobility are prone to biases which under- and mis-represent the travels of different groups of individuals and communities.
As mobility data is gaining prominence within different research communities and practitioners, including the extensive global use during the COVID-19 pandemic, where it is actively used to inform governmental measures and evaluate their effects, the identification and mitigation of these biases is a pressing manner.

In our study of mobile phone mobility datasets from 3 countries we find large imbalances in how much data each wealth group generates: The wealthiest $20\%$ of users contribute more than half of all trips in the dataset, while the poorest $20\%$ account for less than $5\%$ of all trips --- a finding which is consistent across countries and contexts.
One explanation for this observation could be that wealthier individuals simply travel more, and that mobile phone data captures more of these trips.
Travel survey data for cities in Colombia, however, show that this is not the case; poor individuals travel the same distance as wealthier people~\cite{lotero2016rich}.
Similarly, a big-data analysis for mobility patterns in France has shown that there is no connection between travel distance and socioeconomic status~\cite{pappalardo2015using}.
As such, it is unlikely that differences in travel distances can fully explain the vast imbalances observed in the data, which we believe can mainly be understood when taking data generation into account.

The over-representation of data from wealthy individuals is vital to address because there can be differences in where people travel to and which places they visit, especially between social classes.
As Fig.~\ref{fig:networks_statistics} shows, and as the literature documents~\cite{barbosa2021uncovering}, these differences can be pronounced.
Further, as these structural differences persist when evening out the total number of trips across classes (SI Fig 2), it means that they are not merely caused by imbalances in trip volume but signify different mobility characteristics.
As such, neglecting to account for imbalances in data generation we run the risk of institutionalizing discriminatory practices by using mobility datasets which are skewed towards the travel behavior of predominantly wealthy individuals.
This can have a detrimental effect on our ability to accurately predict the evolution of dynamical systems.

To illustrate the potential hazardous effects biases can have on insights drawn from mobility data we use epidemic simulations --- which have been extensively used to understand, and predict the spread of COVID~\cite{oliver2020mobile,kraemer2020effect, Flaxman2020, galeazzi2020human,bonaccorsi2020economic, schlosser2020covid, chang2020mobility,chinazzi2020effect, Gatto2020, arenas2020mathematical}.
In our study we show that biases embedded in data can have a substantial impact on the pattern of epidemic spread.
Biases can cause a severe over- or underestimation of key epidemic properties such as the severity of the peak, or the arrival time of the epidemic; both are characteristics which governments use for planning and responding to pandemics. 
These effects are especially deceiving because of their heterogeneity, as they can heavily impact certain regions but leave others unaffected, and can even go unnoticed when focusing on national averages, an effect that was previously referred to as the \emph{fallacy of the mean}~\cite{vandemoortele2010taking}.
Our research indicates that remote regions are especially impacted by biases, but more research is required to make clear whether there is, in fact, a connection to the demographic properties of the region.
Disease dynamics are one type of dynamical process, but we expect other types of dynamical processes (traffic prediction, population dynamics, migration studies, etc.) to be equally affected by biased data.


To fully answer the question to which degree the debiasing procedure improves representativeness, one would have to compare the data to the true mobility of the population.
Unfortunately, ground truth data with the necessary scale, breadth, and resolution is currently unavailable---We are now aware of any dataset which can be used for a truthful comparison.
One might argue, that other digital data-sources can be used, but they will inevitably not capture \emph{all} trips in the population, and be affected by the same biases.
As such, our study is limited by the same validation factors as other papers focusing on human mobility. 
Nonetheless, the debiasing methods we present here are reasonable mitigation steps as they are derived from a mechanistic framework of how these biases arise, backed by observational data.
It is important to keep in mind that many dynamic processes which are informed by mobility data are inherently probabilistic, where no single scenario, set of parameters, or input data should be seen as the sole truth~\cite{castro2020turning, juul2020fixed}.
In these situations it is vital to model dynamic processes according to multiple alternative scenarios. 
Mobility datasets transformed with our debiasing method would provide one such scenarios where travels of underrepresented individuals are more fairly included.

The debiasing framework we present aims to alleviate the issue and to increase data representativeness.
Our study can be extended in many ways to further explore the topic and improve current practices.
Here, we focus only on a single socio-demographic variable, wealth, to define population sub-groups.
Other factors are equally important in understanding how bias effects human mobility data.
As such, our analysis would benefit from an intersectional approach across multiple additional characteristics including age, gender, ethnicity, and ability~\cite{crenshaw2017intersectionality}.
An important step to facilitate this would be to include more meta-data about the users in mobility datasets, which we urge researchers and MNOs to consider in privacy preserving ways.
We also only focus on mobility data in this study, but we expect the same biases to be present in almost all ``big data'' datasets of human behavior which are passively captured through digital means.
Overall, we show that a failure to account for data generation biases can have distinct real-world effects and urge researchers and policy makers to take these issues into account.


\section*{Methods \& Materials}
\label{sec:methods-materials}

\subsection*{Shannon entropy}

We examine the Shannon entropy $H_i$ of each district $i$ for a given mobility network $F$, given by \cite{Morzy2017}
\begin{equation}
H_i = -\sum_{j\in \Omega_i} p_{ij} \mathrm{log}(p_{ij})   
\end{equation}
where $\Omega_i$ is the set of districts $j$ that district $i$ is connected to, and $p_{ij}$ is the proportion of trips that go from $i$ to $j$,

\begin{equation}
    p_{ij}=\frac{F_{ij}}{\sum_{j\in \Omega_i} F_{ij}}.
\end{equation}

\subsection*{Clustering coefficient}

We calculate the node-wise clustering coefficient $C_i$, which is a measure for the possible triangles through the node that exist, and thus for the cliquishness or transitivity of the network. We use an extension of the clustering coefficient to weighted networks as proposed in \cite{Saramaki2007, Onnela2005, Fagiolo2007}, which is defined as
\begin{equation}
    C_i = \frac{1}{D_i(D_i-1)}\sum_{jk}\left(w_{ij}w_{jk}w_{ki}\right)^{1/3},
\end{equation}
where the edge weights $w_{ij}$ are normalized by the maximum weight in the network, $w_{ij}=F_{ij}/\mathrm{max}(F)$.

\subsection*{Resampling to account for data generation bias}

We debias the mobility network $F$ with respect to data generation bias by over- or undersampling trips from the quintile networks $F^{(q)}$ that are under- or overrepresented, respectively.
Empirically, we find a stark difference in the number of trips 
$$N^{(q)}=\sum_{i,j=1}^{m} F_{ij}^{(q)}$$
recorded in the quintile mobility networks $F^{(q)}$.
If the data generation rate of users were equal across quintiles, and assuming that users undertake the same amount of trips on average irrespective of wealth (see discussion), we would expect the number of trips per quintile network to be equal, $N' = N/5$, where $N=\sum_{i,j=1}^{m} F_{ij}$ is the total number of trips in the network.
Thus, we create a realization of the resampled quintile networks $F'^{(q)}$ by sampling $N'$ trips from the original quintile networks according to
\begin{equation} \label{eq:stoch_sampling_original}
    F'^{(q)} \sim \mathrm{Multinomial}(N', \widehat{p}^{(q)}).
\end{equation}
Here, $\widehat{p}^{(q)}$ is a probability matrix with $p_{ij}$ being the probability to sample a trip for the connection $i\rightarrow j$. We estimate the probabilities from the frequencies of trips recorded in the mobility dataset,
\begin{equation}
    \widehat{p}_{ij}^{(q)} = f_{ij}^{(q)} = \frac{ F^{(q)}_{ij} } { \sum_{ij} F^{(q)}_{ij}} = F^{(q)}_{ij} / N^{(q)}.
\end{equation}
The full resampled mobility network is then given by adding all quintile networks,
\begin{equation}
    F' = F'^{(1)} + F'^{(2)}+ ... + F'^{(5)},
\end{equation}
and contains an equal amount of $N'$ trips from each quintile. A detailed motivation for this procedure is given in SI section~3.3.
One might wonder why we resample the flows as in Eq.\ref{eq:stoch_sampling_original} instead of rescaling the flows $F_{ij}$ by a constant factor to achieve the desired number of trips.
We argue that the sampling method described here better estimates the true structure of the network, see SI section~3.3.5 for details.

\subsection*{Correction for technology access bias}

For districts $i$ where mobility data is present, but where only a fraction  $c_i<1$ of the total population are users of the MNO, we follow the methodology in \cite{tizzoni2014use} and rescale the resampled flows $F'_{ij}$ to calculate the total flow in the population,
$F''_{ij} = F'_{ij}/c_i$.
In districts with no users ($c_i=0$) and no recorded flow ($F'_{ij}=F'_{ji}=0$) at all, we estimate the flows starting and ending in this district from a gravity-like human mobility model \cite{de2011modelling}.
We fit the gravity model to the mobility network of the poorest quintile for each country, as we deem those mobility patterns to be the most representative for the population in regions with no network coverage.
We then add links $G_{ij}$ from the gravity model to the mobility network for all districts with no in- or outgoing flow, yielding the debiased network $\widehat{F}$.
To avoid drastic changes to the structure of the mobility network we do not add all possible flows but only a fraction of links such that density of the network is preserved (see details in SI sec. 3).

\subsection*{Epidemic Simulation}
\label{sec:epidemic_simulation}

We implement a SIR metapopulation model used in~\cite{schlosser2020covid, tizzoni2014use}. A full definition of the model is given in the SI.
In our simulations we use a reproductive number of $R_0 = 2.5$ and recovery rate $\mu=1/6$ days.
We have chosen these epidemic parameters loosely modelled after the wild type of the SARS-CoV-2 virus, for which meta-reviews estimate the basic reproduction number $\mathcal R_0$ in the range of 2-3 and the infectious period as 4-8 days \cite{alimohamadi2020estimate, park2020systematic}.
The epidemic is started with a seed of $I_0=100$ infecteds in a district $i^*$. 
We run simulations on the original mobility network $F$ and on 10 realizations of the debiased network $\widehat{F}$. For each country and network, we run 10 simulations per district, starting the epidemic in all districts present in the original dataset $F$ in turn.
We use this semi-random seeding (instead of choosing a random district each run) to minimize the variation between the simulations due to a different sampling of the districts, and thus improve the comparability of the results.
The arrival time $t_i$ is defined as the first time when the fraction of infecteds $i(t)$ passes the threshold of $0.1\%$ in district $i$.
The epidemic curves in Fig.~\ref{fig:epi_results}a show the median of $i(t)$ and the area that contains the most central $50\%$ of curves over time, where the centrality of curves is determined according to the method described in \cite{juul2020fixed} using the Python software package \emph{curvestat}.

\section*{Acknowledgments}
We thank UNICEF offices in Sierra Leone, Iraq and DRC for their field support and interest on Big Data for social good and for the questions raised on representativeness of Big Data. In particular we thank the technical support of Shane O'Connor, Khulood Malik, Bilal Al-Kiswani, Atif Khurshid, Rhawaa Khalid, Gibson Riungu and Tajudeen Oyewale. We thank Elisa Omodei, Alex Rutherford and Michael Szell for helpful comments on the manuscript. UNICEF Innovation wants to thank Takeda for their Investment in Innovation and Frontier Technology for better health through UNICEF Venture Fund and MagicBox initiatives.

%

\bibliographystyle{unsrt}  
\bibliography{references}  




\end{document}


\section{Datasets}

\subsection{Mobility datasets}
Through partnerships with different Mobile Networks Operators (MNO) we have access to mobility data for Iraq, Sierra Leone, and the Democratic Republic of the Congo (DRC).
Data is captured by one mobile network operator for each country.
The mobility information is extracted from Call Detail Records (CDR). CDRs are originally collected by MNOs for billing purposes, however, because they contain information about which antenna a mobile device is connected to (when making/receiving calls and texts) are frequently used to estimate mobility~\cite{gonzalez2008understanding}.


Due to MNOs having different data science environments there are slight variations in how the trips are calculated from the CDR data in each country.
For Iraq, users are identified by their SIM ids and are first localized to their most frequent tower location, which is denoted as the home location.
This is a common methodology to determine the home location in mobile phone data \cite{pappalardo2021evaluation}.
A related technique is to use only night time locations, but the MNO found that home locations were robust with respect to being inferred only from activity in the 8pm - 8am time-window.
Trips are calculated on an individual level and defined as movements between the home location and all visited antennas.
For instance, if a person has home location \textit{A} and makes calls from antennas in the following pattern: \textit{AABC}, 4 trips will be recorded, \textit{A-A}, \textit{A-A}, \textit{A-B} and \textit{A-C}.

For Sierra Leone and DRC, a slightly different procedure was used to extract trips from the CDR data.
Instead of calculating home locations, trips here are defined as antenna transitions between consecutive calls and texts.
For instance if a person makes calls from antennas: \textit{AABC}, 3 trips will be inferred, \textit{A-A}, \textit{A-B} and \textit{B-C}.
For all datasets, users had to have at least registered two activities (phone calls or text messages) to be included in the mobility datasets.

The datasets are aggregated temporally and geographically to preserve the privacy of users \cite{de2018privacy}.
The aggregation was performed by the MNOs in-house and the aggregate data shared with us.
Spatially, the trips were aggregated to higher administrative regions.
Individuals trips are originally collected on the level of cell towers, which were mapped to the respective administrative regions they are located in. We use the definition of second administrative level boundaries \cite{UNAdminLevels2021}, where we use administrative level 3 for Sierra Leone (corresponding to chiefdoms) and Iraq (corresponding to subdistricts) and administrative level 2 for DRC (corresponding to territories). 
The resulting mobility datasets contain the total number of trips (aggregated over all users) between these spatial regions.
Temporally, all datasets were aggregated over the full time frame of data collection, resulting in one dataset per country. For Iraq, data is collected in May of 2014, in Sierra Leone data is captured from May 1st to December 31st 2015, and in DRC data is captured for a two week period in Dec 2019 - Jan 2020 (1 week in start Dec 2019 and 1 week from start Jan 2020).

In addition, the MNOs provided us with sub-divisions of the mobility datasets which are distinguished by users' income.
Each MNO sorted their users by a proxy metric for their income, and separated them into 5 equal-sized groups (quintiles), ranging from low-income users (Q1) to high-income users (Q5).
Different proxy metrics for income were used by the MNOs. In Iraq and DRC, users were sorted by their airtime expenditure, which is the total amount the users spent for mobile phone activities (making phone calls, sending texts).
In Sierra Leone, users were sorted by the amount of money they spent on topping up their phone account. 
Some general statistics of the full datasets and the quintile networks are given in section~\ref{sec:general_statistics_quintiles}.

\subsection{Geographical and population data}

The mobility datasets are spatially aggregated on the level of administrative level boundaries. We download the corresponding shapefiles for each country from the OCHA Humanitarian Data Exchange \cite{OCHA_HDX}. We use the administrative level 3 data for Sierra Leone and Iraq and level 2 for DRC.

We use population data to calculate several statistics such as the coverage rate $c_i$. The population data was downloaded from Worldpop \cite{worldpop_population} for each country for the year when data was collected.
The population is encoded in raster files which were mapped to the spatial resolution using the above shapefiles, such that the population in each administrative region could be computed.

\section{Structural differences between wealth-disaggregated mobility networks}

\subsection{General statistics}
\label{sec:general_statistics_quintiles}

Here we list some general statistics of the mobility datasets $F$ and their income-quintile disaggregates $F^{(q)}$, in addition to the information given in the main text. The total number of trips recorded in the datasets $F$ is quite similar across countries, despite being recorded over different time frames and with different amounts of users, averaging around 1 billion trips ($1.09e+09$ for Sierra Leone, $8.74e+08$ for DRC, $1.27e+09$ for Iraq). As discussed in the main text, the distribution of trips among the quintile networks $F^{(q)}$ differs considerably though, see Table~\ref{tab:quintile_trip_counts}. The richest quintile Q5 accounts for roughly half of all trips, while the poorest quintile Q1 contains less than $5\%$ of trips.

\begin{table}[tb]
\centering
\begin{tabular}{lrrr}
\toprule
Quintile &        SL &         DRC &         IQ \\
\midrule
1 &   46,835,844 (4.3\%) &   32,540,674 (3.7\%) &   35,681,292 (2.8\%) \\
2 &   76,063,288 (7.0\%)&   58,336,256 (6.7\%)&   98,669,193 (7.8\%) \\
3 &  131,639,612 (12.1\%)&   97,377,816 (11.1\%)&  189,206,744  (14.9\%)\\
4 &  243,054,115 (22.4\%)&  176,319,876 (20.2\%)&  309,134,339  (24.4\%)\\
5 &  588,657,653 (54.2\%)&  508,938,471 (58.3\%)&  633,640,045  (50.0\%)\\
\bottomrule
\end{tabular}
\vspace{1em}
\caption{Number of trips recorded within each income quintile network $F^{(q)}$.}
\label{tab:quintile_trip_counts}
\end{table}

For the number of nodes, we find that only a subset of all administrative regions are present in the datasets (i.e. there was at least one trip recorded as starting in the region). The networks $F$ contain data for 88 regions in Sierra Leone, 132 regions in DRC and 356 in Iraq (see also the comparison to the full dataset in section~\ref{sec:comparison_orig_debiased}). We do not find differences between the quintiles however: If a district is present in one quintile, it is present in all quintiles. This might indicate that, if a district is absent in the data, it is most likely because the respective MNO does not have users in this region at all.

\begin{table}[ht]
\centering
\begin{tabular}{llll}
\toprule
Quintile &            SL &              DRC &             IQ \\
\midrule
1 &    452 (5.90\%) &  1,916 (11.08\%) &  46,675 (38.65\%) \\
2 &    687 (8.97\%) &  2,006 (11.60\%) &  58,809 (48.70\%) \\
3 &   977 (12.76\%) &  2,068 (11.96\%) &  66,694 (54.91\%) \\
4 &  1,275 (16.65\%) &  2,175 (12.58\%) &  71,651 (59.34\%) \\
5 &  1,610 (21.03\%) &  2,296 (13.28\%) &  76,479 (62.25\%) \\
\bottomrule
\end{tabular}
\vspace{1em}
\caption{Number of unique links, and the density of the networks.}
\label{tab:network_density}
\end{table}

The number of links increases for higher quintiles, as does the density of the networks, see Table~\ref{tab:network_density}. There are interesting differences in how the trips are distributed across connections in the quintiles, see Fig.~\ref{fig:network_statistics_distance}. The number of total trips grows with income, but these additional trips are concentrated at the high-flow connections in the network (Fig.~\ref{fig:network_statistics_distance}a). In addition, the higher-income quintiles have more connections over long distances (Fig.~\ref{fig:network_statistics_distance}b). We find this relation for all countries, but the effect is more pronounced in Sierra Leone, less so in DRC and Iraq.

\begin{figure}[htb]
\centering
\includegraphics[width=0.9\linewidth]{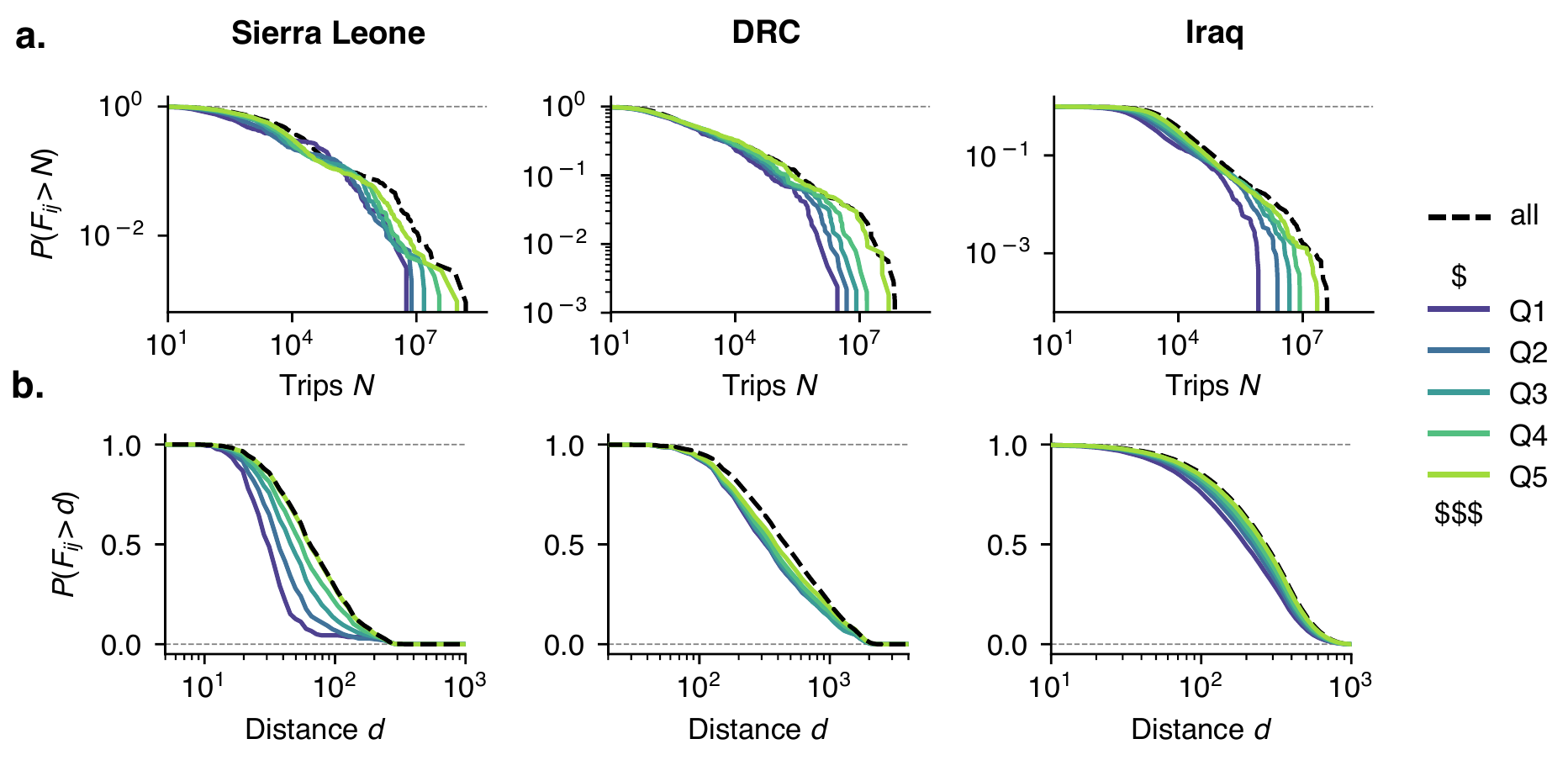}
\caption{Network metrics for the quintile networks $F^{(q)}$. \textbf{a.} The probability distribution of edges $F_{ij}$ having more than $N$ trips. In the richer quintiles such as $Q5$, there are relatively more connections with high flow, indicating that it is especially the high-flow connections in the network that are represented. \textbf{b.} The probability of an edge $F_{ij}$ spanning a distance $d$ or smaller. Low-income quintiles have more connections at smaller distances. The distance of each edge was calculated between the centroids of the connected regions.}
\label{fig:network_statistics_distance}
\end{figure}

Overall, we find that the higher-income mobility networks have more total trips, more unique connections, more high-flow connections and more connections over long distances compared to the low-income networks.

\subsection{Shannon entropy $H_i$}

As detailed in the main text, we observe distinct differences in the node-wise Shannon entropy $H_i$ between the mobility networks $F^{(q)}$ of different income quintiles $q$.
For each country and for each quintile network mobility network $F^{(q)}$, we computed the Shannon entropy $H_i$ of all districts $i=1,\ldots,m$, yielding a set of entropies
$$\mathcal{H}^{(q)}=\{ H_1, \ldots, H_m \}$$ for each quintile $q$.

We performed a two-sample, two-sided Kolmogorov-Smirnov test to determine whether the distributions of the sets $\mathcal{H}^{(q)}$ are statistically significantly different between quintiles. We found that the distributions $\mathcal{H}^{(1)}$ for quintile 1 and $\mathcal{H}^{(5)}$ for quintile 5 are different from each other with high significance for all countries, that is for Sierra Leone ($p=2.1\times10^{-16}$), DRC ($p=1.7\times10^{-7}$) and Iraq ($p=3.3\times10^{-16}$)

\subsection{Clustering coefficient $C_i$}

Similar to the Shannon entropy, we find differences in the weighted clustering coefficient $C$ between the quintile mobility networks $F^{(q)}$.
We calculate the clustering coefficients for all districts $i$ in each quintile $q$, $C^{(q)}=\{C_i\}$.
We compare the sets $C^{(1)}$ and $C^{(5)}$ using a two-sample KS test and find that they are significantly different for all datasets, for Sierra Leone ($p=3.0\times10^{-5}$), DRC ($p=9.4\times10^{-6}$) and Iraq ($p=3.3\times10^{-16}$)

\subsection{Network metrics for resampled networks}

We tested whether the observed differences in the network metrics between the quintiles can be explained solely by a higher amount of total trips in the networks for the richer quintiles. To this end, we calculated the metrics for the resampled networks $F'^{(q)}$, which all contain the same total number of trips (see full description in sec.~\ref{sec:correcting_data_generation_bias}). We found no distinct difference in the distributions (see Fig.~\ref{fig:network_statistics_normalized}) compared to the results for the original networks $F^{(q)}$ (see Figure 2 main text).

\begin{figure}[htb]
\centering
\includegraphics[width=0.9\linewidth]{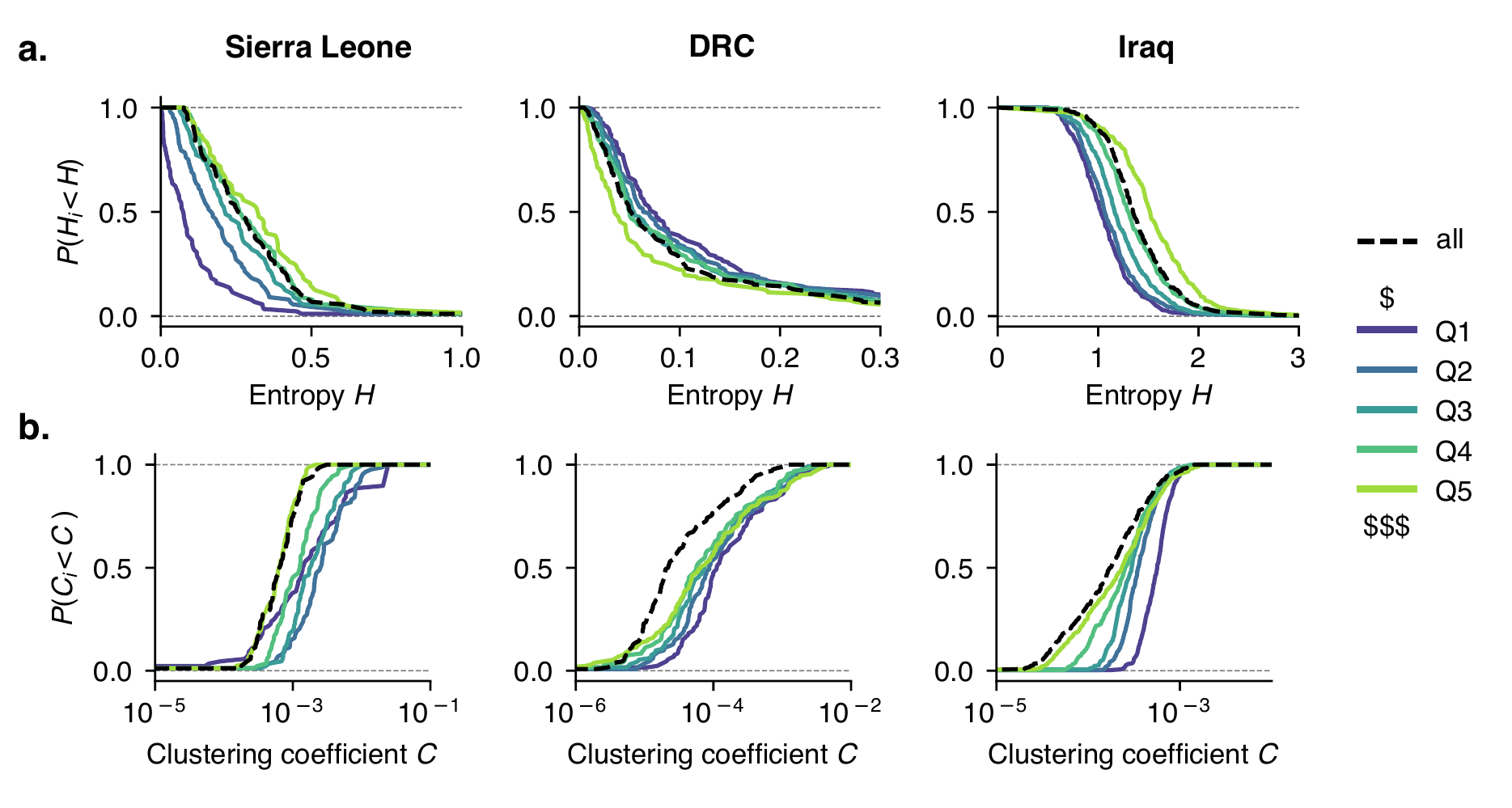}
\caption{Network metrics for the resampled quintile networks $F'^{(q)}$. The resampled networks contain the same amount of total flow in all quintiles. The resulting distributions of the \textbf{a.} Shannon entropy $H_i$ and the \textbf{b.}  weighted clustering coefficient show now distinct difference compared to the results for the original networks (see main text Figure 2).}
\label{fig:network_statistics_normalized}
\end{figure}

\section{Debiasing procedure (Note: Working on content in this part)}

\subsection{Overview}

In the following sections, we describe the full details of the debiasing procedures. We start with a original mobility network $F$, which we assume to be biased with respect to \emph{technology access} and 
\emph{data generation} bias. In the debiasing procedure, we transform the original network in multiple steps:
\begin{itemize}
    \item We account for \emph{data generation} bias, that is the overrepresentation of users based on their data generation rate, by resampling trips from the wealth quintile networks $F^{(q)}$, such that the resulting network contains an equal number of trips from each quintile. The result is the resampled network  $F'$. The procedure is described in sec.~\ref{sec:correcting_data_generation_bias}.
    \item We account for \emph{technology access},
    esampling trips from the quintile networks $F^{(q)}$, such that the resulting network $F'$ contains an equal number of trips from each quintile, see description in sec.~\ref{sec:correcting_data_generation_bias}.
    
\end{itemize}

We have a recorded mobility matrix $F$. Here we will show how we create a debiased mobility matrix $F'$ which is corrected for data generation bias.

\subsection{Correcting for data generation bias}
\label{sec:correcting_data_generation_bias}





\subsubsection{Mobility model}

We assume that a mobility matrix $F$ is generated from trips of individual users in the following way. There are $U$ users which travel among $M$ regions. Each user generates 
\begin{equation}
    n_u = t \cdot c
\end{equation}
trips in the observation timeframe, where $t$ is the actual number of trips a user undertakes, and the collection rate $c$ is the fraction of trips that are recorded in the dataset, with $0 < c\leq 1$.
Thus, the total number of trips generated by users is 
\begin{equation}
    N=U\cdot n_u = U\cdot t \cdot c.
\end{equation}
The probability that a trip generated by a user starts in region $i$ and ends in region $j$ is given as $p_{ij}$ (which are a not-directly measurable property of the system). Then, the expected number of trips along a connection $i \rightarrow j$ is given by
\begin{equation}
    E(F_{ij}) = N \cdot p_{ij} = U\cdot n_u \cdot p_{ij}.
\end{equation}

\subsubsection{Mobility data generating sampling process}

In reality, the number of trips in a recorded datasets $F$ is not equal to the expectation values, i.e. $E(F)$, but it is one possible realisation of a stochastic sampling process:
The flows of the mobility matrix $F$ are drawn from a multinomial distributions, where a total of $N$ trips are distributed according to the probability matrix $p$
\begin{equation} \label{eq:stoch_sampling_original}
    F \sim \mathrm{Multinomial}(N, p).
\end{equation}

\subsubsection{Data generation bias}

What we see from empirical data is that not all users generate the same amount of trips (see main text Fig.~1).
In the income quintiles $q$, each user generates a different number of trips $n_u^{(q)}$, because a different fraction $c^{(q)}$ of actual trips $t$ are recorded in the dataset
\begin{equation}
 n_u^{(q)} = t \cdot c^{(q)}.
\end{equation}
Here we assume that each user undertakes the same amount of real trips $t$ on average, but that the observed differences in the number of trips are explained by differences in the capturing rate $c^{(q)}$ with which trips are recorded, and which varies with income, such that we expect a higher capturing rate for higher income $c^{(1)} < c^{(5)}$.
As a result, although the 5 quintiles contain the same amount of $U/5$ users, they contain a different amount of total trips,
\begin{equation}
    N^{(q)}=U/5\cdot n_u^{(q)},
\end{equation}
where $N^{(1)} < N^{(5)}$.
Moreover, mobility in the quintiles differs in general, i.e. each quintile has different probabilities $p^{(q)}_{ij}$ such that the mobility network is structurally different.
\begin{equation}
    E(F^{(q)}_{ij})=U/5\cdot n_u^{(q)} \cdot p^{(q)}_{ij} = N^{(q)} \cdot p^{(q)}_{ij},
\end{equation}
and 
\begin{equation}\label{eq:resampling_multinomial}
    F^{(q)} \sim \mathrm{Multinomial}(N^{(q)}, p^{(q)}).
\end{equation}

As a result, the aggregate mobility matrix 
\begin{equation}
    F = F^{(1)} + F^{(2)}+ ... + F^{(5)}
\end{equation}
contains an overproportional (or underproportional) amount of trips from quintiles with a higher (lower) capturing rate $b^{(q)}$, and is thus more (less) representative of that quintiles' mobility.

\subsubsection{Correcting for data generation bias}

Here we show how we can correct for the data generation bias outlined in the previous section.
We correct for this bias by estimating the mobility as it would be if each user had the \emph{same} data capturing rate $b^*$, instead of the income-dependent rates $b^{(q)}$.
This means that each user contributes the same number of trips $n_u^*$ to the dataset, which we calculate from the total number of trips $N$ and users $U$ in the dataset, 
\begin{equation}
    n_u^* = N/U = \frac{1}{5}\sum_{q=1}^5 n_u^{(q)} = n_u.
\end{equation}

The true number of trips we would expect in each quintile (if data capturing was uniform among users) is then
\begin{equation}
    N_\mathrm{Q} = N/5
\end{equation}

Then the expected number of trips in each quintile $q$ is
\begin{equation}
    E(F'^{(q)}_{ij})=N_\mathrm{Q} \cdot p_{ij}^{(q)}.
\end{equation}

Thus, when correcting for data generation bias, the expected number of trips changes from the original as

\begin{equation} \label{eq:debias_data_generation_expectation}
    E(F'^{(q)}_{ij})=N_\mathrm{Q} / N^{(q)} \cdot  E(F^{(q)}_{ij}) = w_q \cdot  E(F^{(q)}_{ij}),
\end{equation}

where $w_q = N_\mathrm{Q} / N^{(q)}$ quantifies whether a quintile is over- ($w_q<1$), under- ($w_q>1$) or fairly ($w_q=1$) represented.

\subsubsection{Generating the debiased dataset}

To generate a specific realization of the debiased dataset, we recreate the same stochastic sampling approach that generates the original dataset (see Eq.~\ref{eq:stoch_sampling_original}). For each quintile, we draw that number of trips $N_Q$ that we expect in the debiased dataset,

\begin{equation} \label{eq:stoch_sampling_original}
    F'^{(q)} \sim \mathrm{Multinomial}(N_Q, \hat{p}^{(q)}).
\end{equation}

As we do not know the true probability matrix $p$ with which each connection $i\rightarrow j$ is chosen, we estimate them from the frequencies of trips measured in the recorded dataset,
\begin{equation}
    \widehat{p}_{ij}^{(q)} = f_{ij}^{(q)} = \frac{ F^{(q)}_{ij} } { \sum_{ij} F^{(q)}_{ij}} = F^{(q)}_{ij} / N^{(q)}.
\end{equation}

The final corrected dataset is then given by adding all quintiles,

\begin{equation}
    F' = F'^{(1)} + F'^{(2)}+ ... + F'^{(5)},
\end{equation}

and contains an equal amount of $N_Q$ trips from each quintile.

(Todo: Add pseudo-algorithm that creates debiased dataset)


\subsubsection{Comparison to simple rescaling approach}

One might wonder why we do not simply rescale the mobility flows analogous to Eq.~\ref{eq:debias_data_generation_expectation}, i.e.

\begin{equation} \label{eq:simple_scaling}
    F'^{(q)}_{ij} = w_q \cdot  F^{(q)}_{ij}.
\end{equation}

However, this would not correctly correct the structure of the dataset, and would result in a different network than created by the stochastic sampling process. Specifically, the number of links would not be adapted correctly.

Let's say that a debiased network $F'^{(q)}_{ij}$ contains only 10\% of the trips in the original dataset $F^{(q)}_{ij}$, i.e. $w_q = 0.1$. Intuitively, we expect that the number of links $L$, that is the number of non-zero entries in the mobility matrix, to also be lower than in the original dataset. However, applying the above Eq.~\ref{eq:simple_scaling} would leave the number of links unchanged.

The sampling process outlined above instead does adapt the number of links to the total amount of trips. We have a total of $L_\mathrm{max}=M\times M$ possible links in the system. If we have $N$ trips, which are distributed among individual connections following the probabilities $p_{ij}$, the expected number of links is (see equation 5 in \cite{emigh1983number})
\begin{equation}
    E(L)=L_\mathrm{max}-\sum_{i,j}^M \left\{1-p_{ij}\right\}^N
\end{equation}
which monotonously depends on $N$. Thus, the sampling process accounts for changes in the number of links, while the rescaling approach does not.

\subsection{Correcting for access to technology biases}

Not all individuals in a population are represented in a mobility dataset. 
Access to technology and ownership of mobile phones are unevenly distributed in the population.
There are established methods on how to account for these biases, which we apply here \cite{wesolowski2013impact, tizzoni2014use, pestre2020abcde}.
We use two different approaches to correct for access to technology biases: 
First, for areas where mobility data is present but with biased coverage, we use a normalizing approach to rescale the data.
Secondly, for areas where we have no representation at all in our datasets we estimate mobility flows from a gravity model for human mobility \cite{de2011modelling}.

\subsubsection{Normalizing technology penetration}

For a district $i$, the penetration rate
\begin{equation}
    p_i = \frac{U_i}{P_i}
\end{equation}
quantifies the fraction of users $U_i$ (i.e. the individuals with access to the recording technology) among the total population $P_i$.
Figure~\ref{fig:penetration_scatter}A shows the distribution penetration of the penetration rate $p_i$ in the districts $i$ in our datasets.
As we have $p_i < 1$ in general, not all trips in the population are recorded and enter in our dataset.
In addition, the penetration rate $p_i$ varies across districts, resulting in a under- and over-representation of districts due to different levels of technology access.

We use a simple normalization approach to correct for technology access bias.
The method is applied to each district where we have mobility data, i.e. where $p_i>0$ (the method is referred to as the ``basic normalization'' approach in \cite{tizzoni2014use}):
We estimate the true number of trips $F''_{ij}$ by normalizing the number of trips $F_{ij}$ with the penetration $p_i$ of the district $i$ where the trip originates,

\begin{equation}
    F''_{ij} = 1/p_{i} \cdot F_{ij}
\end{equation}


To demonstrate that the normalized flows $F''_{ij}$ are an improvement over the original flows $F_{ij}$, we compare estimates for the population in the districts using both types of flows. We estimate the population in a district by summing up all flows originating in this district, 

$$ \widehat{P}_i = \frac{\sum_j F_{ij}}{c},$$

where the scale factor $c$ accounts for the number of trips a user does on average in the observation period and is estimated from data,

\begin{equation}
    c = \frac{N}{U},
\end{equation}
with the total number of trips $N=\sum_{ij}F_{ij}$ and the number of users $U$.
Similarly, we estimate the population based on the normalized data,

$$ \widehat{P}''_i = \frac{\sum_j F''_{ij}}{c}.$$

Figure~\ref{fig:penetration_scatter}B shows a comparison of the population estimates $\widehat{P}_i$ and $\widehat{P}_i$ to the census population $P_i$. We find that using the rescaled flows $F''_{ij}$ improves the agreement with census data considerably compared to the original data $F_{ij}$.

\begin{figure}[!htb]
\centering
\includegraphics[width=0.8\linewidth]{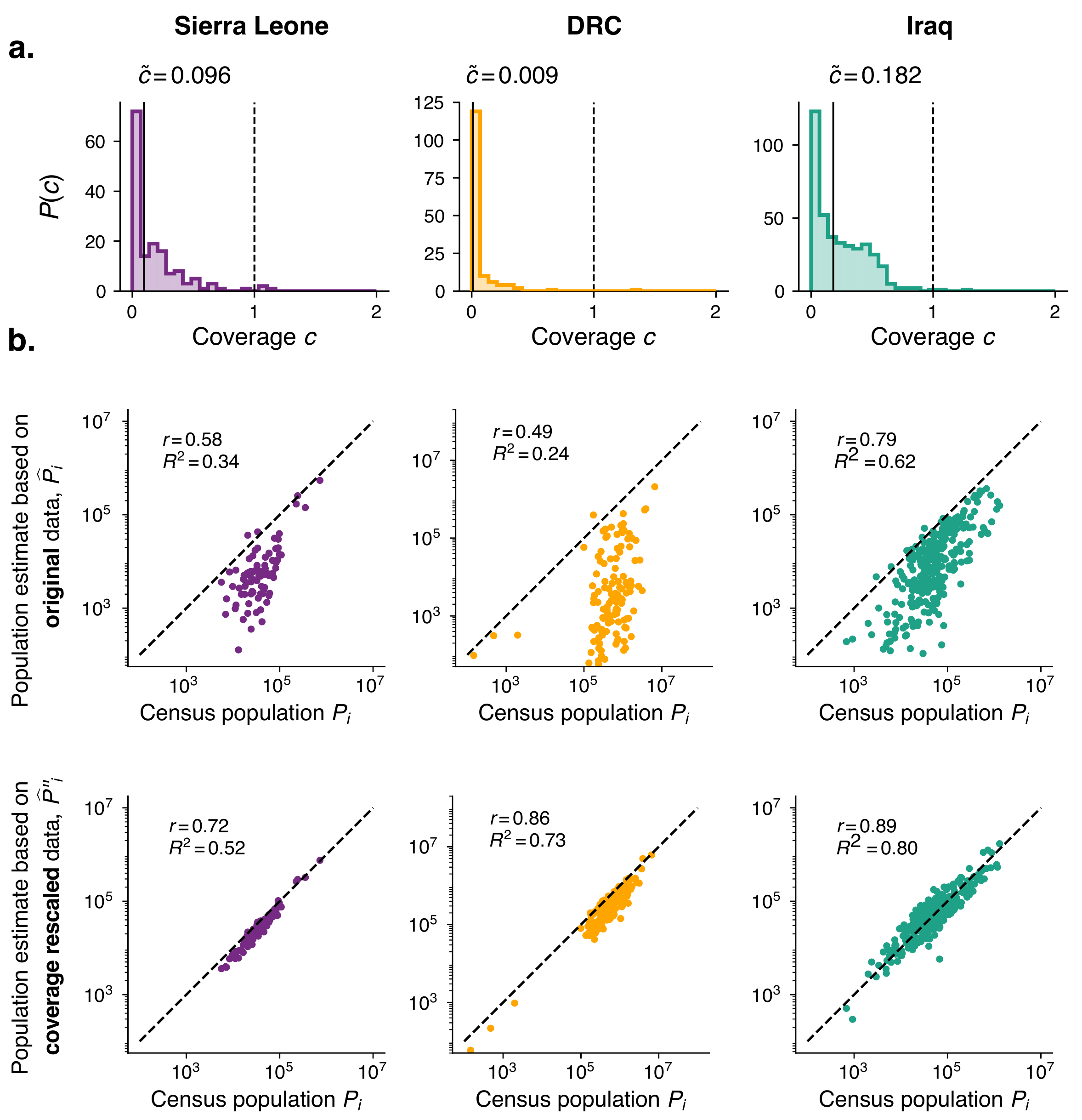}
\caption{\textbf{Variation of technology access.} A) Distribution of the district-wise penetration rates $p_i$ in the datasets. In most districts, only a fraction of the population has access to the technology ($p_i<1$). B) Comparison of population estimates for the original (upper row) and normalized data (lower row) The estimate $\widehat{P}_i$ based on the original data, and (lower row) number of trips originating in each district, compared to the census population $P_i$, before (upper row) and after (lower row) rescaling the flows with the penetration rate $p_i$. Rescaling the flows increases the agreement of estimated and census population considerably.}
\label{fig:penetration_scatter}
\end{figure}

\subsubsection{Estimating missing flows using a gravity model}

For some districts, the datasets contain no recorded mobility at all, i.e. the penetration rate is $p_i=0$.
Districts with no penetration account for $42.5\%$ of all districts in Sierra Leone, $12.0\%$ in DRC and $11.5\%$ in Iraq.
These non-represented districts contain $21.0\%$ of the population in Sierra Leone, $4.2\%$ in DRC and $1.1\%$ in Iraq.
The wide range of these figures show that this type of bias can vary greatly in magnitude, and not accounting for it can potentially leave substantial portion of the population unaccounted for.

For districts with no penetration the simple normalization approach of the previous section fails, as there are no recorded mobility flows $F_{ij}$ that can be scaled. 
Instead, we estimate the missing flows using a theoretical human mobility model.
We use a gravity model, which is often used to model human mobility (CITATIONS).
The gravity flows are given as
\begin{equation} \label{eq:gravity-model}
    G_{ij} = \frac{P_i^\alpha \cdot P_j^\beta}{r_{ij}^\gamma},
\end{equation}
where $P_i$ and $P_j$ are the population in the source and target districts $i$ and $j$, $r_{ij}$ is the distance between the districts (where we use the distance between the centroids of the districts) and $\theta = (\alpha, \beta, \gamma)$ are the parameters of the model.

We fit the gravity model of Eq.~\ref{eq:gravity-model} to the mobility flows we have recorded in our dataset, and then use the resulting parameters to estimate the missing flows. 
We assume that the mobility of quintile 1 best represents the mobility in the districts with missing data ($p_i=0$), as it is likely that these areas that lack coverage have a lower socioeconomic status.
Thus, we fit the gravity model to the flow matrix
\begin{equation}
    F^* = F^{(1)} \cdot N / N^{(1)},
\end{equation}
where $F^*$ are the flows of quintile 1, upscaled such that they sum up to the  same amount of trips as the whole population, $N$.

For a given set of parameters $\theta$, we calculate the error between the gravity model estimates $\hat{G}$ and the target flows $F^*$ using the mean-square-log-error (as we find that the mobility flows $F_{ij}$ are log-normally distributed),
\begin{equation}
    \mathrm{MSLE}(\widehat{G}, F^*) = \frac{1}{|\mathcal{L}|} \sum_{(i,j)\in\mathcal{L}} \left( \mathrm{ln}(1+F_{ij}^*) - \mathrm{ln}(1+\hat{G}_{ij}) \right),
\end{equation}
where $\mathcal{L}$ is the set of all indices $(i,j)$ with non-zero entries in $F^*$. We compute the error using the function \verb|mean_squared_log_error| from the Python package \verb|sklearn.metric|.

We determine the optimal parameters $\theta^*$ using the following procedure: For given parameters $\theta'$, we calculate the mobility flows $\widehat{G}_{ij}$ as well as the error $\mathrm{MSLE}(G,F^*)$, and optimize the parameters $\theta'$ to reduce the error (using the package \verb|scipy.optimize|). The results of the fit are shown in Table~\ref{tab:gravity_fit_params}.

\begin{table}[htb]
\centering
\begin{tabular}{|c || c | c | c | c | c | c |} 
 \hline
  & \multicolumn{3}{|c|}{Original matrix $F$} & \multicolumn{3}{|c|}{Pre-processed matrix $F''$}\\
 \hline
 Country & $\alpha$ & $\beta$ & $\gamma$ & $\alpha$ & $\beta$ & $\gamma$ \\
 \hline
 Sierra Leone & 0.70 & 0.69 & 1.78 & 0.43 & 0.39 & 1.01 \\
 DRC & 0.52 & 0.46 & 1.48 & 0.34 & 0.51 & 1.27 \\
 Iraq & 0.47 & 0.56 & 1.08 & 0.32 & 0.34 & 0.69 \\
 \hline
\end{tabular}
\caption{Optimal parameters $\theta^*$ of the gravity model after fitting. We show both the fit to the original mobility data $F$, and to the matrix $F''$ which has already been resampled and normalized as part of the debiasing (see section~\ref{sec:debiasing_pipeline} for a full description of how the debiasing steps are applied).}
\label{tab:gravity_fit_params}
\end{table}

One quantity that cannot be estimated using the gravity model is the number of intra-district trips $F^*_{ii}$, that is the amount of trips starting and ending in the same district, because $r_{ii}=0$. We estimate these flows in the following way: We calculate the average ratio of intra- to inter-district trips among all districts $m=1,...,M$,
\begin{equation}
    c_1 = \frac{1}{M} \sum_{m=1}^M \frac{F^*_{mm}}{\sum_j F^*_{mj}}.
\end{equation}
Then, after estimating the inter-district flows $\widehat{G}_{mj}$ from the gravity model, we use this empirical ratio $c_1$ to estimate the intra-district flow,
\begin{equation}
    \widehat{G}_{mm} = c_1 \cdot \sum^M_{ \substack{j=1 \\ j\neq m}} \widehat{G}_{mj}
\end{equation}

Finally, we add the estimated gravity flows to the original mobility matrix $F$ to create the imputed network $F'''$. Let $\mathcal{I}$ be the set of all districts $i$ with missing mobility ($p_i=0$) in the original data $F$, that is with no in- and outgoing trips, $F_{ij}=F_{ji}=0$. For these districts $i\in\mathcal{I}$, we add all estimated flows $\widehat{G}_{ij}$ and $\widehat{G}_{ji}$ to the mobility matrix $F$, where we apply an additional thresholding step and only add links above a certain threshold, $\widehat{G}_{ij} > G_c$. The thresholding is applied to preserve the density $\rho_\mathrm{orig}$ of the mobility matrix. For example, the original mobility network $F$ for Sierra Leone contains 88 districts, with $\num{1658}$ links and a density of $\rho=21.6\%$. If we would all possible links $\widehat{G}_{ij}$ for the missing 65 districts $\mathcal{I}$ to and from all districts, we would add $\num{19890}$ links, increasing the density to $\rho=92\%$. To avoid this significant structural change, we add only links greater than the threshold $\widehat{G}_{ij} > G_c$, where $G_c$ is set such that the density of the resulting network $F'''$ is the same as in the original network $F$, that is $\rho''' = \rho_\mathrm{orig}$.

Table~\ref{tab:gravity_network_statistics} shows how the statistics of the network change in the additional gravity flows are taken into account.

\subsection{Creating the debiased dataset}
\label{sec:debiasing_pipeline}

The final debiased dataset is created by merging the results from the debiasing steps. 
Starting with the original mobility dataset $F$, we resample the network to account for data generation bias and create the resampled dataset $F'$. Then, we normalize the flows with the penetration rate to account for technology access bias,
\begin{equation}
    F'' = 1/p_i \cdot F'.
\end{equation}
Finally we add the estimated mobility flows $G$ from the gravity model to create the final debiased dataset,
\begin{equation}
    F_\mathrm{debiased} = F'' + G.
\end{equation}


\section{Comparison of original and debiased networks}
\label{sec:comparison_orig_debiased}

The debiased networks $\widehat{F}$ differ from the original mobility network in many ways $F$. Fig.~\ref{fig:networks_orig_vs_debiased} show the original and debiased networks for Sierra Leone, DRC and Iraq, and Table~\ref{tab:debiasing_statistics} includes some key statistics of the networks. 
Most notably from a visible inspection, the debiased networks contain data for all districts in the country, while the original dataset left regions unrepresented. The number of nodes or regions increases in all districts, as does the population that is represented in the networks (for the latter we count the population that is living in the district). Similarly, the number of total trips in the network increases notably in all districts. This is mainly caused by the upscaling of flows with the coverage rate $c_i$ outlined above. The original dataset only represents a small fraction $c_i<1$ of the population, and upscaling the data to the full population increases the numbers considerably. In addition, adding trips from the gravity model increases the total flow in the network, although the effect is smaller, especially in DRC and Iraq where only a smaller number of districts were not present at all in the original data.

The absolute number of links increases for Sierra Leone but decreases for DRC and Iraq. In the debiasing process, the number of links in general decreases due to the resampling process given in Eq.~\ref{eq:resampling_multinomial}, as not all links present in the original network will necessarily be sampled, is unaffected by the rescaling process, and is increased by the imputation of flows from the gravity model, so that overall it can decrease as well as increase.

However, more interesting is the change in the distribution of flow among the links and nodes in the network, see Fig.~\ref{fig:networks_orig_vs_debiased_statistics}. Overall, the distribution of flow $F$ per link is shifted to higher values, as the total number of trips in the network increases due to debiasing. But the distribution is also less wide and has a smaller relative variance, as indicated by the smaller coefficient of variation $c_v$ in the debiased networks. This is true for both the distribution of flows $F_{ij}$ and node strengths $s_{i}=\sum_j F_{ij}$. This change indicates that the number of trips is more evenly distributed among all possible connections as well as the districts.

\begin{figure}[tb]
\centering
\includegraphics[width=0.7\linewidth]{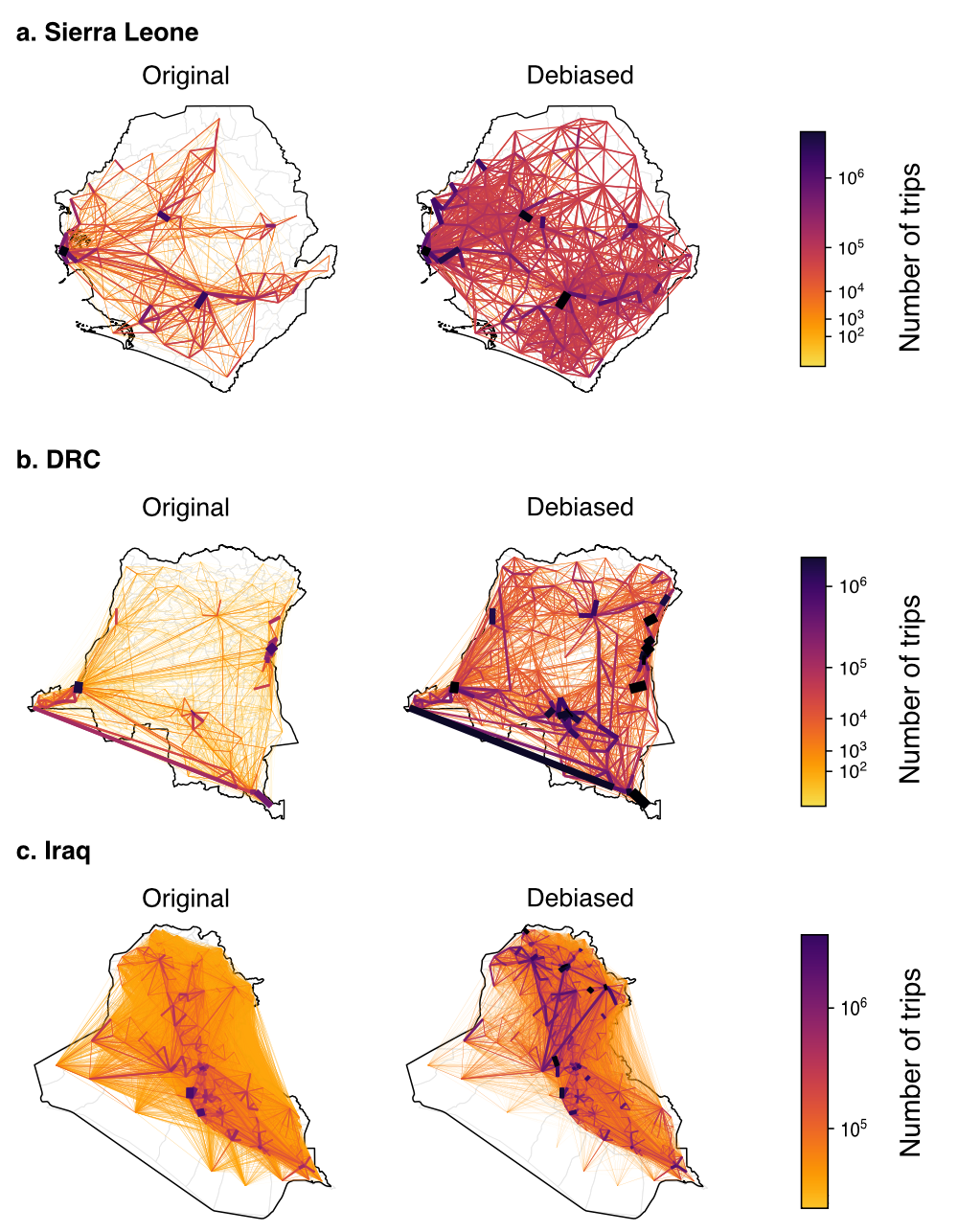}
\caption{\textbf{The original mobility networks compared to the debiased networks.} All networks show marked structural differences due to debiasing. Most visible is the addition of flows from the gravity models, which is less pronounced for the Iraq network as the original network is already very dense. Also visible is the rescaling of flows. In the original network, the flows are centered at cities or densely populated areas, while the network is more decentralized and evenly distributed geographically in the debiased network. The color and width of links is proportional to the flow along each connection, where the scales of both color and widths are tuned for each country separately to make difference more apparent.}
\label{fig:networks_orig_vs_debiased}
\end{figure}

 \begin{table}[htb] 
\centering
\begin{tabular}{|c || c | c | c | c | c | c |} 
 \hline
 \rule{0pt}{3ex}
  & \multicolumn{3}{c|}{Nodes/Regions} & \multicolumn{3}{c|}{Links} \\
 \hline
\rule{0pt}{3ex}
 Country & $F$ & $\widehat{F}$ & $\%$ & $F$ & $\widehat{F}$ & $\%$ \\
 \hline
 \rule{0pt}{3ex}
 Sierra Leone & 88 & 153 & +73\% & 1,658 & 2,807 & +69\%\\
 DRC & 132 & 150 & +13\% & 3,304 & 2,700 & -18\% \\
 Iraq & 352 & 393 & +12\% & 90,242 & 75,393 & -17\% \\
\hline
\hline
 \rule{0pt}{3ex}
 & \multicolumn{3}{c|}{Trips} & \multicolumn{3}{c|}{Population considered regions} \\
 \hline
 \rule{0pt}{3ex}
  Country & $F$ & $\widehat{F}$ & $\%$ & $F$ & $\widehat{F}$ & $\%$ \\
  \hline
  \rule{0pt}{3ex}
  Sierra Leone & 1.09e+09 & 6.98e+09 & + 542\% & 4,864,735 & 6,158,361 & + 26.6\% \\
  DRC & 8.74e+08 & 8.33e+09 & + 853\%  & 104,156,723 & 108,664,385 & + 4.3\% \\
  Iraq &  1.27e+09 & 4.60e+09 & + 263\% & 33,610,757 & 33,970,498 & + 1.07\% \\
 \hline
\end{tabular}
\vspace{1em}
\caption{\textbf{Statistical changes in the network due debiasing}. We compare the several key statistics of the original mobility network $F$ with a realization of the debiased network $\widehat{F}$. Compared are the number of nodes and links in the network, the total number of trips, and the population in all districts that are represented in the network.}
\label{tab:debiasing_statistics}
\end{table}

\begin{figure}[!htb]
\centering
\includegraphics[width=0.9\linewidth]{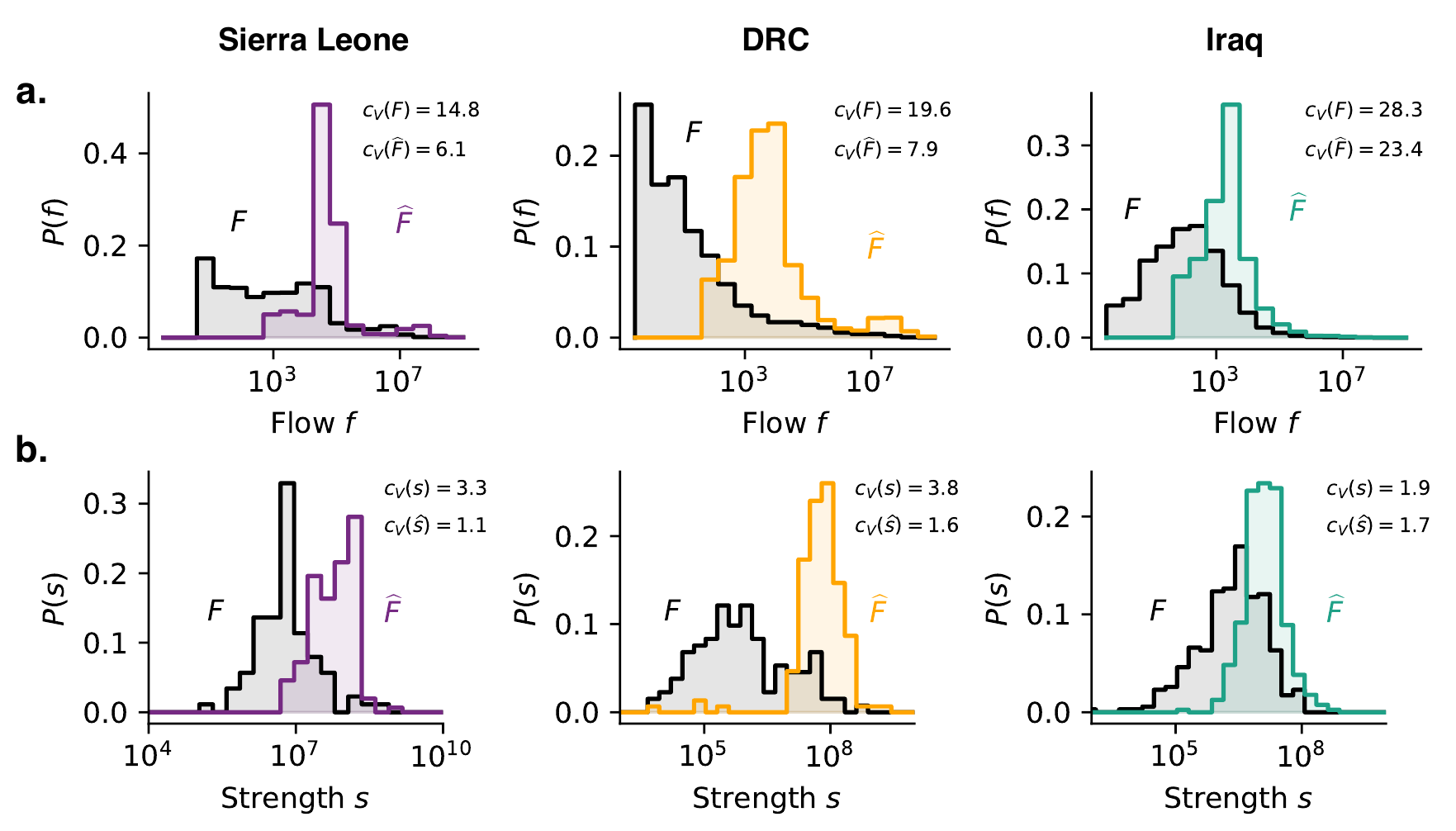}
\caption{\textbf{Differences between the original and debiased networks.} Shown are \textbf{a)} the distribution of flows $f$, that is the number of trips along edges $F_{ij}$, and \textbf{b)} the node strength $s$ in the network, that is the number of outgoing trips from a district, $s_i=\sum_j F_{ij}$. Each distribution is shown for the original network $F$ (black) and the debiased network $\widehat{F}$ (colored). After debiasing, the distributions are shifted to higher values, as the total flow in the network is increased, especially due to the rescaling of flows with the coverage $c_i$. In addition, the values are distributed more closely together, as shown by the decreased coefficient of variation $c_v(x)=\frac{\sigma(x)}{\overline{x}}$, which indicates that the number of trips is more equally distributed among edges and nodes in the mobility network.}
\label{fig:networks_orig_vs_debiased_statistics}
\end{figure}

\section{SIR Simulation}

(ToDo: Include details on how SIR simulation was set up)

We implement a metapopulation SIR-model \cite{Ray1997, Keeling2007} with commuter dynamics based on the model from \cite{Tizzoni2014}. A implementation of the model in Python is available at \url{https://github.com/franksh/EpiCommute}.


\subsection{Epidemic simulation results}

\begin{figure}[htb]
\centering
\includegraphics[width=0.9\linewidth]{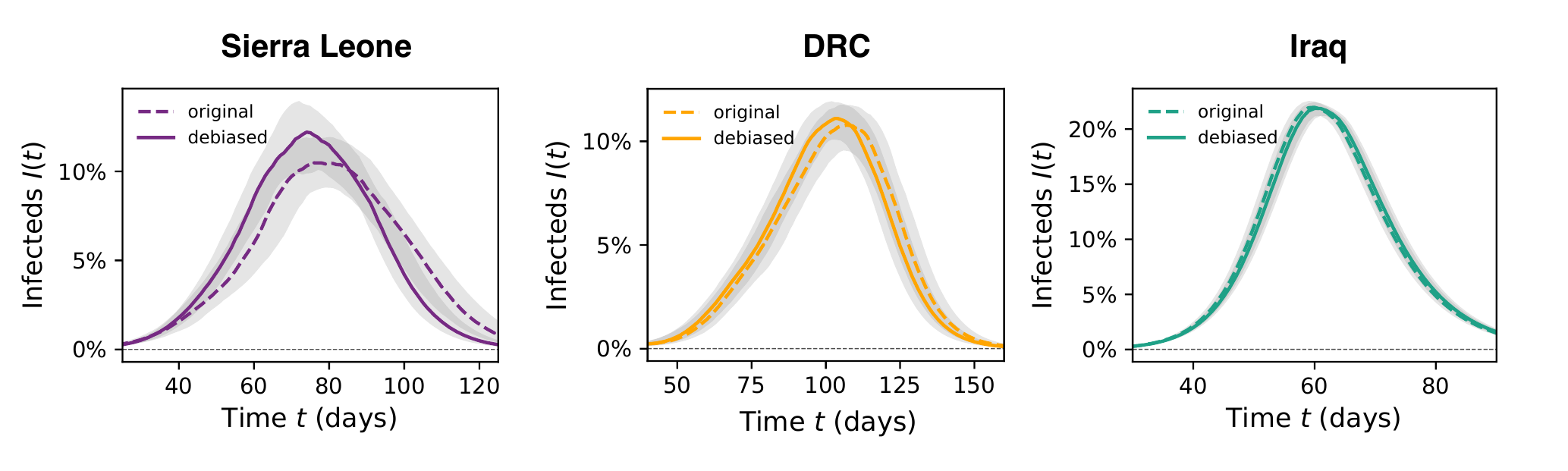}
\caption{\textbf{Additional results of the epidemic simulations.}}
\label{fig:SI_epi_results}
\end{figure}

\section{Internal notes}

\subsection{Arrival time statistics}

This is the arrival statistics from Fig.3 in main manuscript.

\begin{table}[htb]
\centering
\begin{tabular}{lrr}
\toprule
country & Absolute change $\Delta t^\mathrm{abs} = \langle \Delta t^\mathrm{abs}_i \rangle_i$ & Relative change $\Delta t^\mathrm{rel} = \langle \Delta t^\mathrm{rel}_i \rangle_i$ \\
\midrule
DRC &  $-3.26 \pm 1.05 \days$ &  $-3.6 \pm 1.21 \%$  \\
IQ & $-0.84 \pm 0.55 \days$ & $-1.2 \pm 1.63 \%$ \\
SL & $-4.44 \pm 0.74 \days$ & $-6.0 \pm 1.4\%$  \\
SL purple & $+4.62\pm 0.81 \days$ & $+11.8 \pm 2.2 \%$ \\
SL pink & $-3.99 \pm 0.67 \days$ & $-7.8 \pm 1.3 \%$ \\
SL orange & $-26.64 \pm 0.69 \days$ & $-35.7 \pm 0.7\% $ \\
\bottomrule
\end{tabular}
\caption{Arrival time changes, averaged over all districts. Absolute change $\Delta t^\mathrm{abs}_i =  \overline{t^\mathrm{debiased}_i} - \overline{t^\mathrm{original}_i}$ and relative change $\Delta t^\mathrm{rel}_i = (\overline{t^\mathrm{debiased}_i}/ \overline{t^\mathrm{original}_i})-1$ in each district $i$. (Bars denote the mean over the 100 simulations, angular brackets the mean over all districts. Errors are the standard deviation when averaging over simulations and districts.}
\end{table}

\bibliographystyle{unsrt}  
\bibliography{references}


\section{Datasets}

\subsection{Mobility datasets}
Through partnerships with different Mobile Networks Operators (MNO) we have access to mobility data for Iraq, Sierra Leone, and the Democratic Republic of the Congo (DRC).
Data is captured by one mobile network operator for each country.
The mobility information is extracted from Call Detail Records (CDR). CDRs are originally collected by MNOs for billing purposes, however, because they contain information about which antenna a mobile device is connected to (when making/receiving calls and texts) are frequently used to estimate mobility~\cite{gonzalez2008understanding}.


Due to MNOs having different data science environments there are slight variations in how the trips are calculated from the CDR data in each country.
For Iraq, users are identified by their SIM ids and are first localized to their most frequent tower location, which is denoted as the home location.
This is a common methodology to determine the home location in mobile phone data \cite{pappalardo2021evaluation}.
A related technique is to use only night time locations, but the MNO found that home locations were robust with respect to being inferred only from activity in the 8pm - 8am time-window.
Trips are calculated on an individual level and defined as movements between the home location and all visited antennas.
For instance, if a person has home location \textit{A} and makes calls from antennas in the following pattern: \textit{AABC}, 4 trips will be recorded, \textit{A-A}, \textit{A-A}, \textit{A-B} and \textit{A-C}.

For Sierra Leone and DRC, a slightly different procedure was used to extract trips from the CDR data.
Instead of calculating home locations, trips here are defined as antenna transitions between consecutive calls and texts.
For instance if a person makes calls from antennas: \textit{AABC}, 3 trips will be inferred, \textit{A-A}, \textit{A-B} and \textit{B-C}.
For all datasets, users had to have at least registered two activities (phone calls or text messages) to be included in the mobility datasets.

The datasets are aggregated temporally and geographically to preserve the privacy of users \cite{de2018privacy}.
The aggregation was performed by the MNOs in-house and the aggregate data shared with us.
Spatially, the trips were aggregated to higher administrative regions.
Individuals trips are originally collected on the level of cell towers, which were mapped to the respective administrative regions they are located in. We use the definition of second administrative level boundaries \cite{UNAdminLevels2021}, where we use administrative level 3 for Sierra Leone (corresponding to chiefdoms) and Iraq (corresponding to subdistricts) and administrative level 2 for DRC (corresponding to territories). 
The resulting mobility datasets contain the total number of trips (aggregated over all users) between these spatial regions.
Temporally, all datasets were aggregated over the full time frame of data collection, resulting in one dataset per country. For Iraq, data is collected in May of 2014, in Sierra Leone data is captured from May 1st to December 31st 2015, and in DRC data is captured for a two week period in Dec 2019 - Jan 2020 (1 week in start Dec 2019 and 1 week from start Jan 2020).

In addition, the MNOs provided us with sub-divisions of the mobility datasets which are distinguished by users' wealth.
Each MNO sorted their users by a proxy metric for their wealth, and separated them into 5 equal-sized groups (quintiles), ranging from low-wealth users (Q1) to high-wealth users (Q5).
Different proxy metrics for wealth were used by the MNOs. In Iraq and DRC, users were sorted by their airtime expenditure, which is the total amount the users spend on mobile phone activities (making phone calls, sending texts).
In Sierra Leone, users were sorted by the amount of money they spent on topping up their phone account.
(At the time of data collection unlimited calling and texting plans were not available in any of the countries.)
In DRC the MNO did not link their payment database to the CRD data, instead airtime expenditure was estimated by weighting call-minutes to SMS according to a 3:1 ratio, such that sending one text message is 1/3 the cost of making a 1 minute call. 
Some general statistics of the full datasets and the quintile networks are given in section~\ref{sec:general_statistics_quintiles}.

\subsection{Geographical and population data}

The mobility datasets are spatially aggregated on the level of administrative level boundaries. We download the corresponding shapefiles for each country from the OCHA Humanitarian Data Exchange \cite{OCHA_HDX}. We use the administrative level 3 data for Sierra Leone and Iraq and level 2 for DRC.

We use population data to calculate several statistics such as the coverage rate $c_i$. The population data was downloaded from Worldpop \cite{worldpop_population} for each country for the year when data was collected.
The population is encoded in raster files which were mapped to the spatial resolution using the above shapefiles, such that the population in each administrative region could be computed.

\section{Structural differences between wealth-disaggregated mobility networks}

\subsection{General statistics}
\label{sec:general_statistics_quintiles}

Here we list some general statistics of the mobility datasets $F$ and their wealth-quintile disaggregates $F^{(q)}$, in addition to the information given in the main text. The total number of trips recorded in the datasets $F$ is quite similar across countries, despite being recorded over different time frames and with different amounts of users, averaging around 1 billion trips ($1.09e+09$ for Sierra Leone, $8.74e+08$ for DRC, $1.27e+09$ for Iraq). As discussed in the main text, the distribution of trips among the quintile networks $F^{(q)}$ differs considerably though, see Table~\ref{tab:quintile_trip_counts}. The richest quintile Q5 accounts for roughly half of all trips, while the poorest quintile Q1 contains less than $5\%$ of trips.

\begin{table}[tb]
\centering
\begin{tabular}{lrrr}
\toprule
Quintile &        SL &         DRC &         IQ \\
\midrule
1 &   46,835,844 (4.3\%) &   32,540,674 (3.7\%) &   35,681,292 (2.8\%) \\
2 &   76,063,288 (7.0\%)&   58,336,256 (6.7\%)&   98,669,193 (7.8\%) \\
3 &  131,639,612 (12.1\%)&   97,377,816 (11.1\%)&  189,206,744  (14.9\%)\\
4 &  243,054,115 (22.4\%)&  176,319,876 (20.2\%)&  309,134,339  (24.4\%)\\
5 &  588,657,653 (54.2\%)&  508,938,471 (58.3\%)&  633,640,045  (50.0\%)\\
\bottomrule
\end{tabular}
\vspace{1em}
\caption{Number of trips recorded within each wealth quintile network $F^{(q)}$.}
\label{tab:quintile_trip_counts}
\end{table}

For the number of nodes, we find that only a subset of all administrative regions are present in the datasets (i.e. there was at least one trip recorded as starting in the region). The networks $F$ contain data for 88 regions in Sierra Leone, 132 regions in DRC and 356 in Iraq (see also the comparison to the full dataset in section~\ref{sec:comparison_orig_debiased}). We do not find differences between the quintiles however: If a district is present in one quintile, it is present in all quintiles. This might indicate that, if a district is absent in the data, it is most likely because the respective MNO does not have users in this region at all.

\begin{table}[ht]
\centering
\begin{tabular}{llll}
\toprule
Quintile &            SL &              DRC &             IQ \\
\midrule
1 &    452 (5.90\%) &  1,916 (11.08\%) &  46,675 (38.65\%) \\
2 &    687 (8.97\%) &  2,006 (11.60\%) &  58,809 (48.70\%) \\
3 &   977 (12.76\%) &  2,068 (11.96\%) &  66,694 (54.91\%) \\
4 &  1,275 (16.65\%) &  2,175 (12.58\%) &  71,651 (59.34\%) \\
5 &  1,610 (21.03\%) &  2,296 (13.28\%) &  76,479 (62.25\%) \\
\bottomrule
\end{tabular}
\vspace{1em}
\caption{Number of unique links, and the density of the networks.}
\label{tab:network_density}
\end{table}

The number of links increases for higher quintiles, as does the density of the networks, see Table~\ref{tab:network_density}. There are interesting differences in how the trips are distributed across connections in the quintiles, see Fig.~\ref{fig:network_statistics_distance}. The number of total trips grows with wealth, but these additional trips are concentrated at the high-flow connections in the network (Fig.~\ref{fig:network_statistics_distance}a). In addition, the higher-wealth quintiles have more connections over long distances (Fig.~\ref{fig:network_statistics_distance}b). We find this relation for all countries, but the effect is more pronounced in Sierra Leone, less so in DRC and Iraq.

\begin{figure}[htb]
\centering
\includegraphics[width=0.9\linewidth]{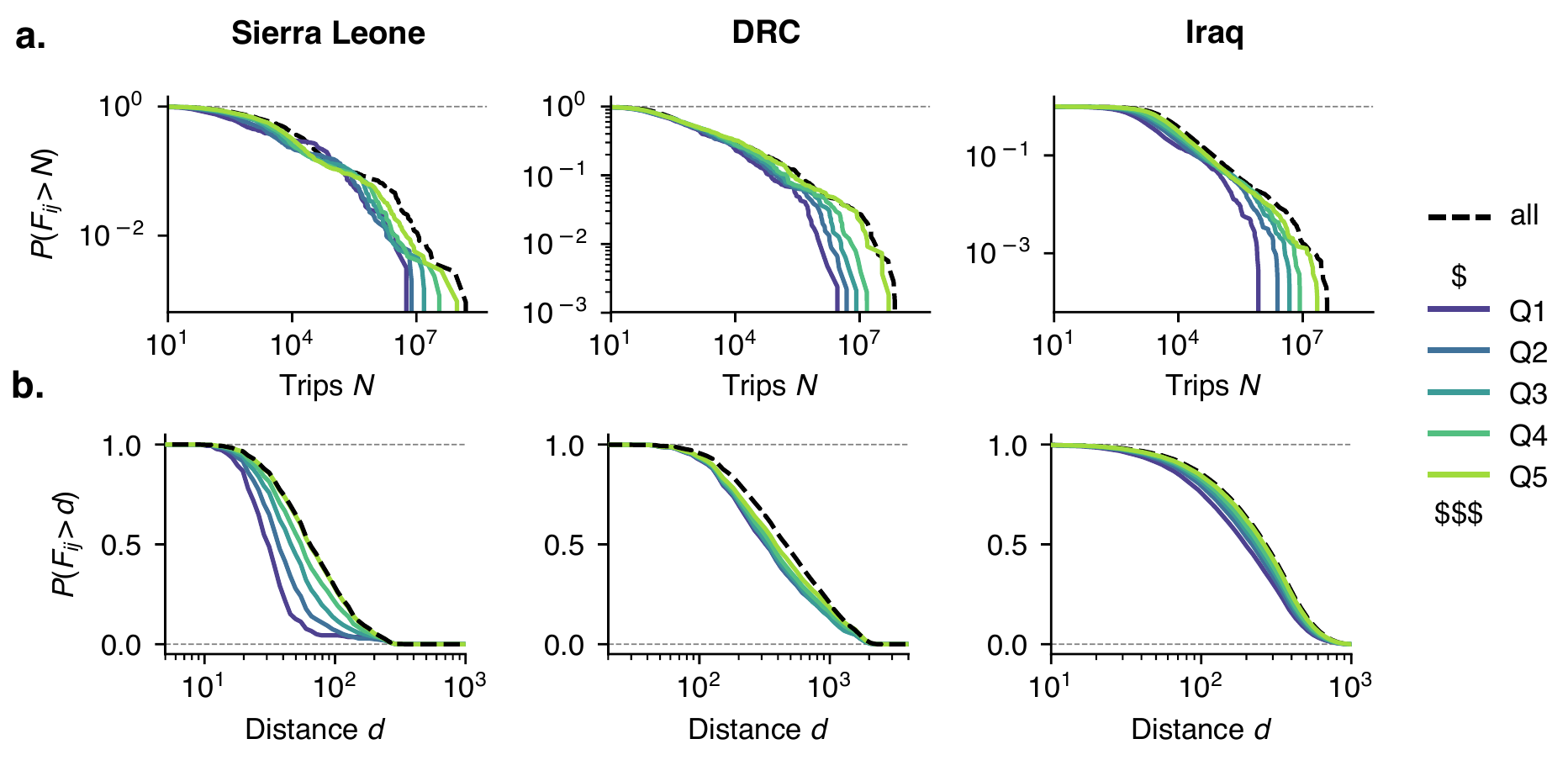}
\caption{\textbf{Network metrics for the quintile networks $F^{(q)}$}. \textbf{a.} The probability distribution of edges $F_{ij}$ having more than $N$ trips. In the richer quintiles such as $Q5$, there are relatively more connections with high flow, indicating that it is especially the high-flow connections in the network that are represented. \textbf{b.} The probability of an edge $F_{ij}$ spanning a distance $d$ or smaller. Low-wealth quintiles have more connections at smaller distances. The distance of each edge was calculated between the centroids of the connected regions.}
\label{fig:network_statistics_distance}
\end{figure}

Overall, we find that the higher-wealth mobility networks have more total trips, more unique connections, more high-flow connections and more connections over long distances compared to the low-income networks.

\subsection{Shannon entropy $H_i$}

As detailed in the main text, we observe distinct differences in the node-wise Shannon entropy $H_i$ between the mobility networks $F^{(q)}$ of different wealth quintiles $q$.
For each country and for each quintile network mobility network $F^{(q)}$, we computed the Shannon entropy $H_i$ of all districts $i=1,\ldots,m$, yielding a set of entropies
$$\mathcal{H}^{(q)}=\{ H_1, \ldots, H_m \}$$ for each quintile $q$.

We performed a two-sample, two-sided Kolmogorov-Smirnov test to determine whether the distributions of the sets $\mathcal{H}^{(q)}$ are statistically significantly different between quintiles. We found that the distributions $\mathcal{H}^{(1)}$ for quintile 1 and $\mathcal{H}^{(5)}$ for quintile 5 are different from each other with high significance for all countries, that is for Sierra Leone ($p=2.1\times10^{-16}$), DRC ($p=1.7\times10^{-7}$) and Iraq ($p=3.3\times10^{-16}$)

\subsection{Clustering coefficient $C_i$}

Similar to the Shannon entropy, we find differences in the weighted clustering coefficient $C$ between the quintile mobility networks $F^{(q)}$.
We calculate the clustering coefficients for all districts $i$ in each quintile $q$, $C^{(q)}=\{C_i\}$.
We compare the sets $C^{(1)}$ and $C^{(5)}$ using a two-sample KS test and find that they are significantly different for all datasets, for Sierra Leone ($p=3.0\times10^{-5}$), DRC ($p=9.4\times10^{-6}$) and Iraq ($p=3.3\times10^{-16}$)

\subsection{Network metrics for resampled networks}

We tested whether the observed differences in the network metrics between the quintiles can be explained solely by a higher amount of total trips in the networks for the richer quintiles. To this end, we calculated the metrics for the resampled networks $F'^{(q)}$, which all contain the same total number of trips (see full description in sec.~\ref{sec:correcting_data_generation_bias}). We found no distinct difference in the distributions (see Fig.~\ref{fig:network_statistics_normalized}) compared to the results for the original networks $F^{(q)}$ (see Figure 2 main text).

\begin{figure}[htb]
\centering
\includegraphics[width=0.9\linewidth]{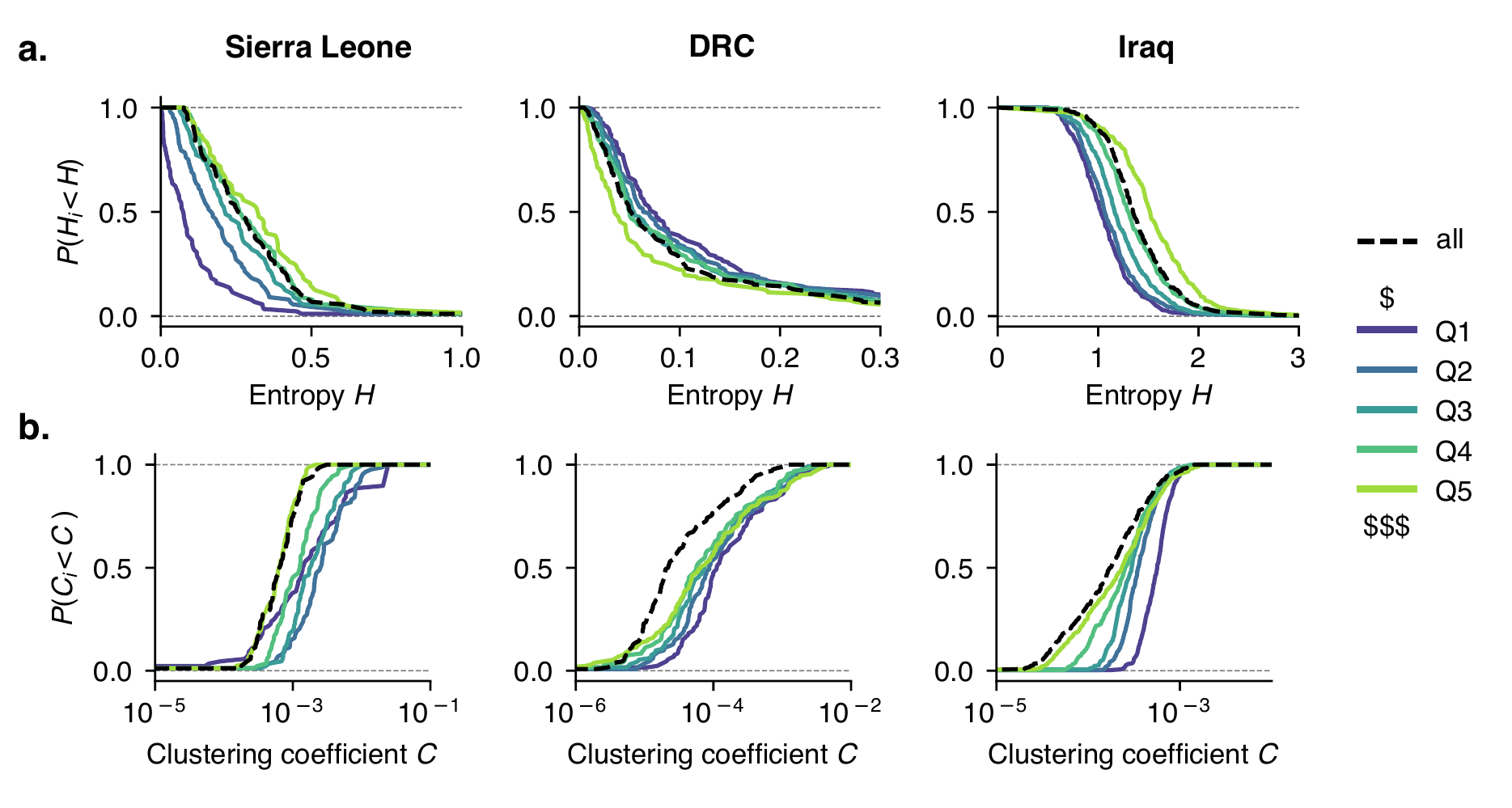}
\caption{\textbf{Network metrics for the resampled quintile networks $F'^{(q)}$}. The resampled networks contain the same amount of total flow in all quintiles. The resulting distributions of the \textbf{a.} Shannon entropy $H_i$ and the \textbf{b.} weighted clustering coefficient show no distinct differences compared to the results for the original networks (see main text Figure 2), indicating that the observed discrepancies between the quintiles in both metrics are not caused by a different total flow.}
\label{fig:network_statistics_normalized}
\end{figure}

\section{Debiasing mobility data}

\subsection{Overview}

In the following sections, we describe the full details of the debiasing procedures, which we outline here. We start with a given empirical mobility network $F$. The network $F$ is a sample of the underlying true mobility network $F^*$ which includes all movements by all individuals in the population. The aim of the debiasing procedures is to create an estimate $\widehat{F}$, that takes into account various types of biases, and is a better approximation of the true mobility network $F^*$ than the network $F$.

The procedure used to debias the data and create the estimate $\widehat{F}$ is outlined in section~\ref{sec:debiasing_procedure}. Before, however, we want to formulate a general model for data generation to motivate the debiasing procedure used here. Specifically, we formulate a general model for the expected number of trips \E{\user{\Fij}} one individual $u$ in the population contributes to the network $F$. We introduce the \emph{technology access} and \emph{data generation} bias as explicit parameters in this general model, and detail the assumptions about those parameters we make in this manuscript which simplify the general model.

The specific model for individual mobility can then be used to counter the effects of biases in the dataset. In the debiasing procedure, we transform the original network in multiple steps:
\begin{itemize}
    \item We account for \emph{data generation} bias, that is the overrepresentation of users based on their data generation rate, by resampling trips from the wealth quintile networks $F^{(q)}$, such that the resulting network contains an equal number of trips from each quintile. The result is the resampled network  $F'$. The procedure is described in sec.~\ref{sec:correcting_data_generation_bias}.
    \item We account for \emph{technology access} bias, that is the fact that only a fraction of the population is present in in the datasets, by rescaling the original flows to account for the missing individuals and by estimating flows from a gravity model for areas where no data is present at all in the dataset. The procedures are described in sections \ref{sec:debiasing_technology_access} and \ref{sec:gravity_model}.
\end{itemize}

If the reader is purely interested in the debiasing steps applied here, they may skip to sec.~\ref{sec:debiasing_procedure}, while the full framework is given in the next sections.

\subsection{Mathematical framework for debiasing}

\subsubsection{Mobility networks as a realization of a stochastic sampling process}

In order to adequately debias the mobility network $F$, we have to make assumptions as to how the network $F$ is generated. In particular, we here assume that the network $F$ is a realization of a stochastic sampling process which we outline below.

We assume that there is an underlying true mobility network  $F^*$, consisting of all trips of all individuals in the population undertaken in the time frame of data collection. The mobility network is measured for a certain spatial tessellation $S$. If we observe a random trip, it is not equally likely to occur among all possible connections $i\rightarrow j$ between regions, as individuals are more likely to travel along some routes than others, population densities vary, etc. We assume that there are (unknown) underlying probabilities $p_{ij}$ that a trip goes from region $s_i$ to region $s_j$.

If we measure trips in the population using some recording technology, we obtain a mobilty dataset $F$, which is a sample of the true mobility network $F^*$. Let us say that $F$ consists of $N$ trips. For now, we assume that each trip in $F^*$ has an equal probability to be recorded the network $F$. This also means that there are no differences between the individuals in the population, and that each individuals' mobility follows the same probabilities $p_{ij}$ (in the next sections we model how biases can affect the sampling procedure). Then, $F$ is generated by sampling $N$ trips from the probabilities $p$ using a multinomial sampling process,

\begin{equation}\label{eq:general_multinomial}
F\sim\mathrm{Multinomial}(N,p),    
\end{equation}

meaning that the probability to observe $n$ trips along a connection $i \rightarrow j$ is given by

\begin{equation}
    P\left(F_{ij}=n\right)= {N\choose k} p_{ij} \left(1-p_{ij}\right)^{N-n}.
\end{equation}
 
This means the that expected value of number of trips along an edge is given as 

\begin{equation}\label{eq:expectation_Fij_general}
  \E{\Fij}=N\cdot p_{ij}.
\end{equation}

In the following, we will often limit our discussion to the expectation values for brevity, but we always refer to them in relation to this underlying sampling process.
 
 
 
In general, a measured mobility network $F$ is only an approximation of the underlying mobility network $F^*$. One obvious reason is that the network $F$ is only a sample of the true mobility and would change between measurements. But more importantly, the equality-assumptions detailed above are likely not true in practice. The way in which individuals are sampled can differ, affecting their representation in the dataset or their likelihood to be included at all. In addition, the mobility of individuals, that is the probabilities $p_{ij}$, can differ.

Biases are likely to enter mobility datasets already on the level of individual data collection. Thus, in the next section we formulate a general model of individual data generation, which accounts for different sources of biases. We then specify the general model with the assumptions made in this publication.

 \subsubsection{General individual data generation model}

In this section we formulate a general model of individual data generation, following the approach from the previous section, but focusing on the mobility network of a single individual.

Let $u$ be a single individual from the population of the area depicted by $F^*$. Note that we do not yet assume that $u$ is a user of the recording technology---the individual $u$ is present in the true mobility network $F^*$, but not necessarily in the recorded dataset $F$. Let $F^{*(u)}$ be the true mobility network of the individual $u$ during the given time frame, that is the full collection of trips the person undertook.

We are interested in the mobility network $\user{F}$, that is the actual recorded mobility network of the individual in the dataset $F$. Aggregating these individual mobility networks $\user{F}$ then yields the complete mobility network, $F = \sum_u \user{F}$.

We assume that the individual mobility network $\user{F}$ is sampled according to a multinomial sampling procedure, similar to the previous section. However, the specifics of the sampling now depend on the properties of the individual $u$.
Say that the individual $u$ is defined by a set of features $\left\{ x_k^{(u)} \right\}_{k=1 \ldots K}$. These can for example stand for socioeconomic properties such as wealth, education level, housing situation, etc., or general information about the individual such as age or gender---any property that might reasonably affect how an individual interacts with the recording technology. In specific contexts, relevant features that are deemed most influential have to be identified and singled out, such as we do here later.

We first state the mathematical form of our model for the individual mobility network $\user{F}$, give a brief description of its components and then go into more details about the parameters later on.
Say we are recording all trips in a population of $P$ individuals using a specific recording technology. We are recording over a time period $T$ measured in days.
Then, the expected number of trips between two regions $i$ and $j$ that any individual $u$ in the population contributes to the dataset can be given as
\begin{equation}\label{eq:general_data_generation_model}
    \E{\user{\Fij}} = \user{c}_{ij} \cdot \user{g}_{ij} \cdot r \cdot T \cdot \user{p}_{ij},
\end{equation}
where
\begin{itemize}
    \item $\user{c}_{ij}$ is the \emph{technology coverage probability}, $0\leq \user{c}_{ij}\leq 1$, which is the probability that the individual $u$ has access to the recording technology, i.e. is a user of the technology (for example mobile phones).
    \item $\user{g}_{ij}$ is the \emph{data generation rate}, which is the rate at which a user produces data activities per time unit (for example per day). In our mobile phone data sets, these activities are text messages or phone calls.
    \item $r$ is the recording probability, that is the probability that a data activity leads to one trip being added to the dataset.
    \item $T$ is the total duration of data collection in time units.
    \item and $\user{p}_{ij}$ is the probability that the captured trip is recorded along the connection $i \rightarrow j$.
\end{itemize}

Combining all these factors yields the expected number of trips that the individual contributes. In this general form, the parameters $c$, $g$ and $p$ can depend on the properties of the regions $s_i$ and $s_j$, the properties of the individual $u$, and other external factors. An application of this framework requires one to make assumptions about the functional form of these parameters, which depends on which biases one deems to be relevant or neglectable, or which information can be inferred from available data. We will apply the model to our case in the following.

\subsubsection{Data generation in the absence of biases and for uniform mobility}

Before we apply the general data generation model to our system, we want to show the case of no biases and homogeneous mobility. First, we assume that there is no technology access bias, i.e. that every individual $u$ in the population has access to the recording technology and that coverage is homogeneous across space, i.e. $\user{c}_{ij}=1$. Second, we assume that each individual generates the same amount of data activities (and independent of location), i.e. $\user{g}_{ij}=g$. Third, we assume that the mobility patterns of individuals are uniform, that they have the same probabilities to undertake a trip along a connection, i.e. $\user{p}_{ij}=p_{ij}$. Then, the expected number of trips per individual is

\begin{equation}
     \E{\user{\Fij}} = g \cdot r \cdot T \cdot p_{ij},
\end{equation}

and for the whole population of $P$ individuals

\begin{equation}
     \E{\Fij} = \sum_{u=1}^P \E{\user{\Fij}} = P \cdot g \cdot r \cdot T \cdot p_{ij} = N \cdot p_{ij},
\end{equation}

where we recover the original formulation given in Eq.~\ref{eq:expectation_Fij_general}.

\subsubsection{Specific individual data generation model}
\label{sec:individual_data_generation_model}

For our study, we make several assumptions about the specific form of Eq.~\ref{eq:general_data_generation_model}, motivated by the factors we want to focus on and by the data we have available to determine parameters.

The coverage $\user{c}_{ij}$ is the probability that an individual $u$ in the population is a user of the recording technology. In its most general form, it can depend on the features $\left\{ x_k^u \right\}_{k=1 \ldots K}$ of the individual (such as wealth, education, age or gender) as well as on the regions $s_i$ and $s_j$ where the trip starts and ends,
\begin{equation}
    \user{c}_{ij} = f\left({x_1^{(u)}, x_2^{(u)}, \ldots,x_K^{(u)}, s_i, s_j  }\right).
\end{equation}
In our study, we have information given about the number of mobile phone users $U_i$ per spatial region $s_i$, which we can use for debiasing. We thus assume that the coverage only depends on the spatial region in which the trip starts,
\begin{equation}
    \user{c}_{ij} = f\left(s_i\right) = c_i.
\end{equation}
The coverage can then be estimated from our data as
\begin{equation}
    \widehat{c}_i = U_i / P_i,
\end{equation}
where $U_i$ are the number of mobile phone users and $P_i$ the population in region $s_i$.

The data generation rate $\user{g}_{ij}$ is the number of data activities an individual produces. Mostly, we expect this to again depend on the properties of the individual, which determine their technology usage patterns. For example, one can expect that young people send more text messages than older individuals.
In our study, we focus on an individuals wealth $x_\mathrm{wealth}^u$ (as measured by the proxy of airtime expenditure) as the sole variable that determines the data generation rate,
\begin{equation}
    \user{g}_{ij} = h\left({x_1^u, x_2^u, \ldots,x_K^u, s_i, s_j  }\right) \approx h'(x_\mathrm{wealth}^u) = \user{g}.
\end{equation}

We modify the general data generation model Eq.~\ref{eq:general_data_generation_model} with the above assumptions regarding the coverage and data generation rate and retrieve

\begin{equation}
\label{eq:invidual_Fij_expectation}
    \E{\user{\Fij}} = c_{i} \cdot \user{g} \cdot r \cdot T \cdot \user{p}_{ij}.
\end{equation}

Finally, in our study we cannot distinguish the mobility of individuals, but we know the aggregate mobility of 5 equal-sized quintiles $q$ of the total user base of $U$ users (corresponding to 5 quintiles of the population $P$). 
This also means that we cannot distinguish the properties of individuals, but only know them on a quintile-level. This concerns the mobility patterns $\q{p}_{ij}$, which we assume to be homogeneous among the individuals of each quintile, as well as the data generation rate \q{g}.
Thus, we can coarsen the model to yield the number of trips in each quintile $q$ as 

\begin{equation}
\label{eq:quintile_Fij_expectation}
    \E{\q{\Fij}} =  P/5 \cdot c_{i} \cdot \q{g} \cdot r \cdot T \cdot \q{p}_{ij}.
\end{equation}

\subsection{Debiasing procedure}
\label{sec:debiasing_procedure}

In this section, we give a detailed description of how we generate a realization of the debiased mobility network $\widehat{F}$. We first generate the network $F'$ by accounting for data generation bias, then create the network $F''$ by account for unequal technology access, and finally impute missing flows from a gravity model, yielding the final network $\widehat{F}$. A brief summary of the methodology is afterwards given in sec.~\ref{sec:debiasing_summary}

\subsubsection{Correcting for data generation bias}
\label{sec:correcting_data_generation_bias}

We first correct for bias stemming from inequalities in the data generation
of individuals. In section 2, we showed that the empirical mobility
networks $\q G$ of richer users contain more trips than those of
poorer users. Using the above framework, we can explain this by differences
in the data generation rate $\q g$ between quintiles.

Our main idea is to correct for data generation bias by resampling
the network, where we sample more trips from quintiles with a lower
data generation rate, and fewer trips from quintiles with a higher
data generation rate. In the following, we derive the procedure for
how to do this from the above data generation framework. To briefly
summarize the result: We generate the corrected quintile flow matrix
$F'^{(q)}$ by sampling $N'=\frac{1}{5}N$ trips (where $N=\sum_{i,j=1}^{M}\q{F_{ij}}$
is the total number of trips in the original quintile network $\q F$)
from the original quintile networks $\q F$, where we sample each
connection $i\rightarrow j$ according its frequency in the original
network, i.e. according to the normalized weights $w_{ij}$.

Let's assume we are given a network $G$ and sub-networks $\q G.$
We understand these networks to be realizations of the multinomial
sampling process defined in \ref{eq:general_multinomial}.
The networks have a varying number of trips $\q N$ per quintile as
the data generation rate $\q g$ varies. We want to compute a corrected
network $G'$ (and corresponding quintile networks $\q{G'}$), which
is the mobility network if each user had the same data generation
rate $g'$. A straightforward choice is to set $g'$ equal to the
average data generation rate in the population,
\[
g'=\frac{1}{5}\sum_{q\in\mathcal{Q}}g^{(q)}.
\]
We find that the number of trips in the corrected quintile networks
$\q{G'}$ is then the same, $N'^{(q)}=N'$, and that it can be calculated
from the data as as

\begin{eqnarray*}
N' & = & \sum_{i,j=1}^{M}\E{F_{ij}^{'(q)}}\\
 & \overset{\mathrm{Eq.}\ref{eq:quintile_Fij_expectation}}{=} & \sum_{i,j=1}^{M}\frac{P}{5}\cdot c_{i}\cdot g'\cdot r\cdot t_{\mathrm{max}}\cdot\q{p_{ij}}\\
 & \approx & \frac{P}{5}\cdot c\cdot g'\cdot r\cdot t_{\mathrm{max}}\underset{=1}{\underbrace{\sum_{i,j=1}^{M}\q{p_{ij}}}}\\
 & = & \frac{P}{5}\cdot c\cdot\left(\frac{1}{5}\sum_{q\in\mathcal{Q}}\q g\right)\cdot r\cdot t_{\mathrm{max}}\\
 & = & \frac{1}{5}\cdot\sum_{q\in\mathcal{Q}}\frac{P}{5}\cdot c\cdot\q g\cdot r\cdot t_{\mathrm{max}}\\
 & = & \frac{1}{5}\cdot\sum_{q\in\mathcal{Q}}\q N\\
 & = & \frac{1}{5}N.
\end{eqnarray*}
The estimated number of trips per quintile, $N'$, in case of an equal
data generation rate $g'$, is thus the average number of trips $\q N$
per quintile, or $1/5$th of the total number of trips in the network
$N$. This result is not suprising, but it is nice that it follows
from our mechanistic data generation model. Here, we used the approximation
that the coverage $c_{i}$ is independent of regions $i$. We correct
for the coverage rate separately in the next section.

We can now generate realizations of the corrected quintile flow matrics
$\q{F'}$ using the multinomial sampling approach:
\[
\q{F'}\sim\mathrm{Multinomial}(N',\q p).
\]
where we estimate the relevant quantities from the recorded flow matrics
$\q F$: The number of trips as derived above, 
\[
N'=\frac{1}{5}N=\frac{1}{5}\sum_{i,j=1}F_{ij}.
\]
The probabilities $\q p$ are unknown, but can be estimated from the
recorded trip frequencies, or normalized weights $w_{ij}$, as
\[
\widehat{p}_{ij}^{(q)}=w_{ij}=\frac{\q{F_{ij}}}{\sum_{i,j=1}^{M}\q{F_{ij}}}.
\]
We generate $F'^{(q)}$ by sampling $N=1/5$ trips from the original
quintile flow matrices $\q F$ according to the weights $w_{ij}$.
The fully corrected mobility network $G'$ is then retrieved by adding
the flow matrices of all resampled quintile networks,
\[
F'=F'^{(1)}+F'^{(2)}+\cdots+F'^{(5)}.
\]
In the corrected network $G'$, the users of each quintile contribute
the same amount of $N'$ trips, and are represented with equal weight
in the full network.

A final note: From the above formulation, it follows that the expected
values of the flows scale as 
\[
\E{F_{ij}^{'(q)}}=\frac{N'}{\q N}\E{\q{F_{ij}}}.
\]
One might then wonder why we go through the trouble of resampling
the network and not simply rescale the flows as
\[
F_{ij}^{'(q)}\approx\frac{N'}{\q N}\q{F_{ij}},
\]
that is by omitting the expectation values. We argue that this approach
is a worse estimate for mobility as it doesn't correctly adjust the
structural properties of the mobility network, that is it doesn't
recreate the characteristics of the network as they would be if they
were measured in reality (as described by the sampling process). This
is especially true if the flows are downsampled, i.e. $N'<\q N.$
We discuss this in more detail in section \ref{sec:error_in_scaling_approach}.

\subsubsection{Correcting for technology access bias}
\label{sec:debiasing_technology_access}

Here, we account for technology access bias, resulting in the rescaled network $F''$. Not all individuals in a population have equal access to the recording technology, which is accounted for by the coverage $c_i$ in our definition of the specific individual data generation model in sec.~\ref{sec:individual_data_generation_model}. Again, we here assume that the coverage only depends on the region of origin of the trip $c_i$. Our approach follows previous studies, which have found that the coverage can vary across spatial regions in mobility datasets \cite{wesolowski2013impact, tizzoni2014use, pestre2020abcde}.

For our datasets, we know the number of users $U_i$ in each districts, which let's us calculate the local coverage
\begin{equation}
    \widehat{c}_i = \frac{U_i}{P_i}.
\end{equation}
We find that the coverage varies widely across districts $i$, and is considerably smaller than 1 for most districts, see Fig.~\ref{fig:penetration_scatter}a.

To correct for technology access bias, we want to estimate the mobility network $F''$ if the coverage were $c_i=c=1$ in all districts, i.e. meaning that all individuals in the population have access to the recording technology. Again, we use the individual data generation model, specifically the expectation value for the user-specific mobility flows $\E{\user{\Fij}}$ given in Eq.~\ref{eq:invidual_Fij_expectation}. By setting $c_i=1$, we find

\begin{equation}
    \E{F''^{(u)}_{ij}} = \user{g} \cdot V \cdot \user{p}_{ij},
\end{equation}

from which follows by comparison to Eq.~\ref{eq:invidual_Fij_expectation}

\begin{equation}
    \E{F''^{(u)}_{ij}} = \frac{1}{c_i} \cdot \E{F^{(u)}_{ij}},
\end{equation}

and analogously for the full network

\begin{equation}
    \E{F''_{ij}} = \frac{1}{c_i} \cdot \E{F_{ij}}.
\end{equation}

To determine the flows $F''_{ij}$ now, based on the expectation values, one could follow the multinomial sampling approach as detailed before. However, in contrast to the data generation bias, we argue that here it is actually a valid approximation to simply rescale the flows by the coverage as
\begin{equation}
\label{eq:rescale_coverage}
    F''_{ij} \approx \frac{1}{c_i} F_{ij},
\end{equation}
because the flows are in general upscaled by a considerable amount, as $1/c_i \gg 1$ for most districts. We explain this in more detail in sec.~\ref{sec:error_in_scaling_approach}.

Finally, we want to demonstrate that the coverage rescaled network $F''$ corresponds better to the census population $P$ than the original network $F$ by constructing a simple estimator for the census population size $P_i$ in each district. First, we calculate the average number of trips per user from the data,
\begin{equation}
    \user{n} = \frac{N}{U},
\end{equation}
where $N=\sum_{ij} F_{ij}$ is the total number of trips in the network and $U$ the number of users. Then, a simple estimate for the population size in a district is
\begin{equation}
    \widehat{P}_i = \frac{N_i}{\user{n}},
\end{equation}
where $N_i$ are all trips originating in the district, $N_i=\sum_{i} F_{ij}$. We can estimate $\widehat{P}_i$ as stated above, but we can also estimate it based on the coverage-rescaled network $F''$,
\begin{equation}
    \widehat{P^{\prime\prime}_i} = \frac{N''_i}{\user{n}}.
\end{equation}
In Fig.~\ref{fig:penetration_scatter}, we compare both estimates and find that the coverage-rescaled network $F''$ provides a considerably better estimate of the population sizes in the districts as given by the census population.

\begin{figure}[!htb]
\centering
\includegraphics[width=0.8\linewidth]{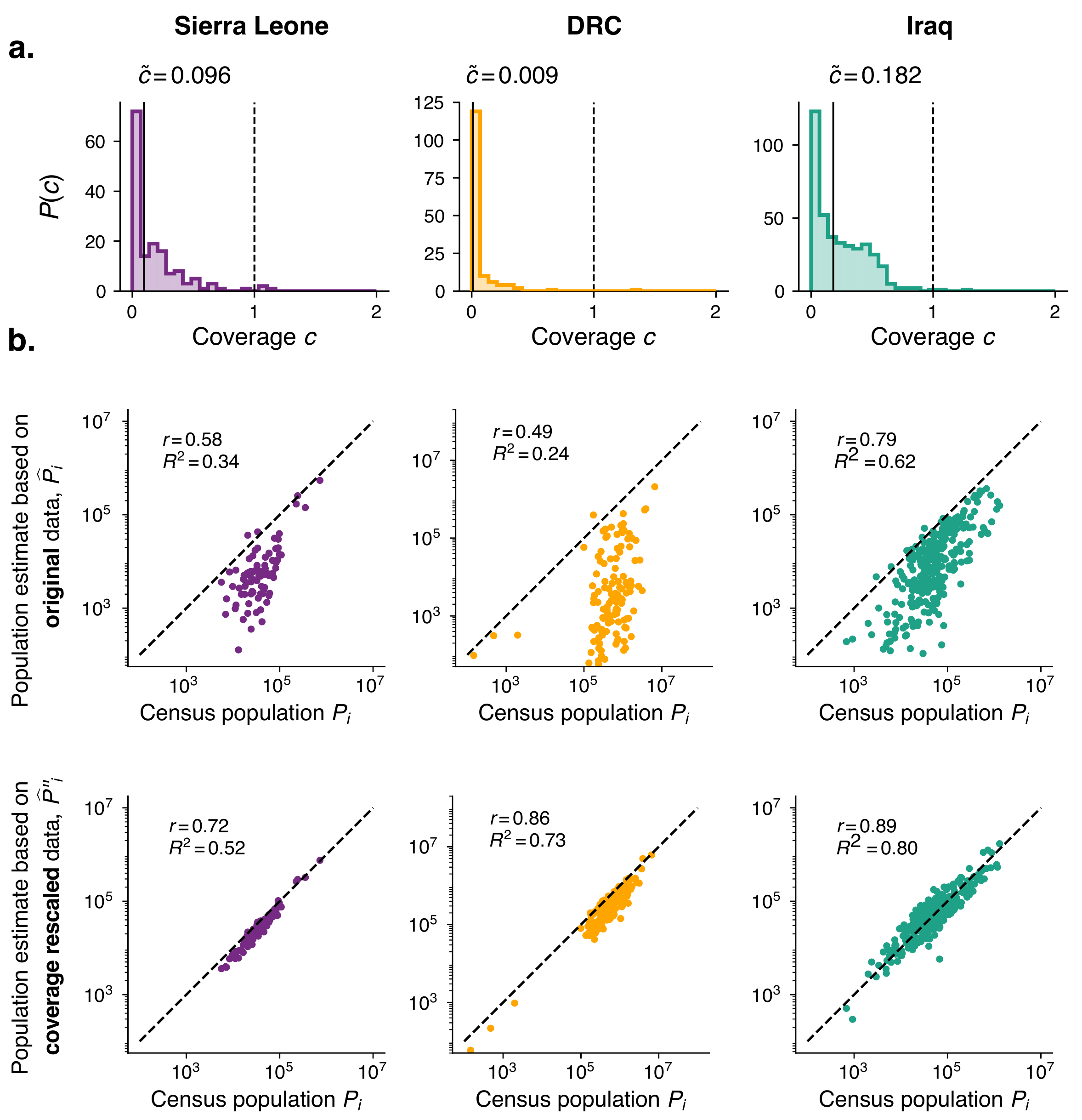}
\caption{\textbf{Variation of technology access.} \textbf{a.} Distribution of the district-wide coverage $c_i$. In most districts, only a fraction of the population has access to the recording technology ($c_i<1$), as seen from the median $\tilde{c}$. For some districts we find $c_i>1$, which indicates an error in either our data of $U_i$ or $P_i$. \textbf{b.} Comparison of population estimates for the original (upper row) and coverage-rescaled data (lower row). The estimate $\widehat{P}_i$ is based on the number of trips originating in each district compared to the census population $P_i$, before (upper row) and after (lower row) rescaling the flows with the coverage $c_i$. Rescaling the flows increases the agreement of estimated and census population considerably.}
\label{fig:penetration_scatter}
\end{figure}

\subsubsection{Estimating missing flows from a gravity model}
\label{sec:gravity_model}

For some districts, the datasets contain no recorded mobility at all, i.e. the coverage is $c_i=0$.
Districts with no coverage account for $42.5\%$ of all districts in Sierra Leone, $12.0\%$ in DRC and $11.5\%$ in Iraq.
These non-represented districts contain $21.0\%$ of the population in Sierra Leone, $4.2\%$ in DRC and $1.1\%$ in Iraq.
The wide range of these figures show that this type of bias can vary greatly in magnitude, and not accounting for it can potentially leave substantial portion of the population unaccounted for.

For districts with no coverage the simple normalization approach of the previous section fails, as there are no recorded mobility flows $F_{ij}$ that can be scaled.
Instead, we estimate the missing flows using a theoretical human mobility model.
We use a gravity model, for which the flows are given as
\begin{equation} \label{eq:gravity-model}
    G_{ij} = \frac{P_i^\alpha \cdot P_j^\beta}{r_{ij}^\gamma},
\end{equation}
where $P_i$ and $P_j$ is the census population in the source and target districts $i$ and $j$, $r_{ij}$ is the distance between the districts (where we use the distance between the centroids of the districts) and $\theta = (\alpha, \beta, \gamma)$ are the parameters of the model.

We fit the gravity model of Eq.~\ref{eq:gravity-model} to the mobility flows we have recorded in our dataset, and then use the resulting parameters to estimate the missing flows. 
We assume that the mobility of quintile 1 best represents the mobility in the districts with missing data ($c_i=0$), as it is likely that these areas that lack coverage have a lower socioeconomic status.
Thus, we fit the gravity model to the flow matrix
\begin{equation}
    F^* = F^{(1)} \cdot N / N^{(1)},
\end{equation}
where $F^{(q)}$ are the flows of quintile 1, upscaled such that they sum up to the same amount of trips as the whole population, $N$.

For a given set of parameters $\theta$, we calculate the error between the gravity model estimates $\widehat{G}$ and the target flows $F^*$ using the mean-square-log-error (as we find that the mobility flows $F_{ij}$ are approximately log-normally distributed),
\begin{equation}
    \mathrm{MSLE}(\widehat{G}, F^*) = \frac{1}{|\mathcal{L}|} \sum_{(i,j)\in\mathcal{L}} \left( \mathrm{ln}(1+F_{ij}^*) - \mathrm{ln}(1+\widehat{G}_{ij}) \right),
\end{equation}
where $\mathcal{L}$ is the set of all indices $(i,j)$ with non-zero entries in $F^*$. We compute the error using the function \verb|mean_squared_log_error| from the Python package \verb|sklearn.metric|.

We determine the optimal parameters $\theta^*$ using the following procedure: For given parameters $\theta'$, we calculate the mobility flows $\widehat{G}_{ij}$ as well as the error $\mathrm{MSLE}(\widehat{G},F^*)$, and optimize the parameters $\theta'$ to reduce the error (using the package \verb|scipy.optimize|). The results of the fit are shown in Table~\ref{tab:gravity_fit_params}.

\begin{table}[htb]
\centering
\begin{tabular}{|c || c | c | c | c | c | c |} 
 \hline
  & \multicolumn{3}{|c|}{Original matrix $F$} & \multicolumn{3}{|c|}{Pre-processed matrix $F''$}\\
 \hline
 Country & $\alpha$ & $\beta$ & $\gamma$ & $\alpha$ & $\beta$ & $\gamma$ \\
 \hline
 Sierra Leone & 0.70 & 0.69 & 1.78 & 0.43 & 0.39 & 1.01 \\
 DRC & 0.52 & 0.46 & 1.48 & 0.34 & 0.51 & 1.27 \\
 Iraq & 0.47 & 0.56 & 1.08 & 0.32 & 0.34 & 0.69 \\
 \hline
\end{tabular}
\caption{Optimal parameters $\theta^*$ of the gravity model after fitting. We show both the fit to the original mobility data $F$, and to the matrix $F''$ which has already been resampled and normalized as part of the debiasing (see previous sections).}
\label{tab:gravity_fit_params}
\end{table}

One quantity that cannot be estimated using the gravity model is the number of intra-district trips $F^*_{ii}$, that is the amount of trips starting and ending in the same district, because $r_{ii}=0$. We estimate these flows in the following way: We calculate the average ratio of intra- to inter-district trips among all districts $m=1,...,M$,
\begin{equation}
    a_1 = \frac{1}{M} \sum_{m=1}^M \frac{F^*_{mm}}{\sum_j F^*_{mj}}.
\end{equation}
Then, after estimating the inter-district flows $\widehat{G}_{mj}$ from the gravity model, we use this empirical ratio $a_1$ to estimate the intra-district flow,
\begin{equation}
    \widehat{G}_{mm} = a_1 \cdot \sum^M_{ \substack{j=1 \\ j\neq m}} \widehat{G}_{mj}
\end{equation}

Finally, we add the estimated gravity flows $\widehat{G}$ to the mobility matrix $F''$ to create the imputed network $F'''$. Let $\mathcal{I}$ be the set of all districts $i$ with missing mobility ($c_i=0$) in the data $F''$, that is with no in- and outgoing trips, $F''_{ij}=F''_{ji}=0$. For these districts $i\in\mathcal{I}$, we add all estimated flows $\widehat{G}_{ij}$ and $\widehat{G}_{ji}$ to the mobility matrix $F''$. 

Here, we apply an additional thresholding step and only add links above a certain threshold, $\widehat{G}_{ij} > G_c$. The thresholding is applied to preserve the density $\rho$ of the mobility matrix. For example, the original mobility network $F$ for Sierra Leone contains 88 districts, with $\num{1658}$ links and a density of $\rho_\mathrm{orig}=21.6\%$. If we would all possible links $\widehat{G}_{ij}$ for the missing 65 districts $\mathcal{I}$ to and from all districts, we would add $\num{19890}$ links, increasing the density to $\rho=92\%$. To avoid this significant structural change, we add only links greater than the threshold $\widehat{G}_{ij} > G_c$, where $G_c$ is set such that the density of the resulting network $F'''$ is the same as in the original network $F''$, that is $\rho'' = \rho'''$ (note that this density can still differ from the original density $\rho_\mathrm{orig}$ as the previous resampling is likely to have changed the density, but as we think that the resampled and rescaled network $F''$ is a better estimate of the true mobility $F^*$ we chose to retain its density).

\subsubsection{Summary of the debiasing procedure}
\label{sec:debiasing_summary}

Here we summarize how we create a realization of the debiased dataset $\widehat{F}$ using the methods outlined in the previous sections. We start with the original mobility network $F$ as given from data, which we are also given disaggregated into the wealth quintile networks $F^{'(q)}$. First, we sample the quintile mobility networks $F^{'(q)}$ under equal data generation by drawing $N'=N/5$ trips from the multinomial distribution Eq.\ref{eq:multinomial_data_generation}),
\begin{equation}
\label{eq:multinomial_data_generation}
    F'^{(q)} \sim \mathrm{Multinomial}\left(N', \q{p}\right)
\end{equation}
with the probabilities
$\widehat{p}^{(q)}_{ij} = \q{F}_{ij} / \sum_{i,j} \q{F}_{ij} $.
The aggregate network $F'$ is retrieved by adding all sampled estimates for the quintiles,
\begin{equation}
    F' = F'^{(1)} + F'^{(2)}+ ... + F'^{(5)}.
\end{equation}

Second, we rescale the flows in $F'$ to account for technology access bias with the measured coverage rates $c_i = U_i/P_i$ as given in Eq.~\ref{eq:rescale_coverage},
\begin{equation}
    F^{\prime\prime}_{ij} \approx \frac{1}{c_i} F'_{ij},
\end{equation}
where $U_i$ is the number of mobile phone users and $P_i$ the population in district $i$.

Third, we impute missing flows for districts with no data ($c_i=0$) from a gravity model estimate $\widehat{G}$ according to the methodology in the previous section. This last steps yielding the final estimate $\widehat{F}$.

\subsubsection{Comparison of the stochastic sampling procedure with a simple multiplicative approach}
\label{sec:error_in_scaling_approach}

Throughout the debiasing procedure we have taken care to point out that we assume that a mobility dataset $F$ is generated from an underlying sampling process as 

\begin{equation}
F\sim\mathrm{Multinomial}(N,p),
\end{equation}

where $N$ is the total number of trips in the network and the probabilities $p_{ij}$ determine how they are distributed among the flows $F_{ij}$. Under this assumption, it is in general not a good approximation to rescale the flows by a multiplicative factor if one wishes to change the number of trips in the network. Say we want to obtain a mobility network $\tilde{F}$ with 
\begin{equation}
    \tilde{N} = a\cdot N,
\end{equation}
trips, where $a>0$. In our framework it follows that

\begin{equation}
    \E{\tilde{F}} = a\cdot \E{F},
\end{equation}
but this does not mean that individual flows can be rescaled as
\begin{equation}
\label{eq:rescaling_wrong}
    \tilde{F}_{ij} \neq a\cdot F_{ij}.
\end{equation}

The flows $\tilde{F}_{ij}$ are a random variable that can change between realizations. If the number of trips $\tilde{N}$ is very high, or if the scaling factor is high, $a\gg1$, we can expect the flows to converge to their expectation value,

\begin{equation}
    \tilde{F}_{ij}\xrightarrow[]{N\rightarrow\infty}\E{\tilde{F}_{ij}}.
\end{equation}

However, if $\tilde{N}$ is small or $a<1$ we can expect the error to be greater.

In particular, a multiplicative rescaling can only increase or decrease the flow, but not account for whether a flow is present at all, which can have great effect on the network structure. For example, let us assume $a=0.1$, so that the rescaled network $\tilde{F}$ contains only $10\%$ of the trips of the original network $F$. Intuitively, we would expect that the number of links $L$ in the mobility network to decrease as well. However, a simple rescaling as in Eq.~\ref{eq:rescaling_wrong} would leave the number of links unchanged.

The multinomial sampling process outlined above instead does adapt the number of links to the total amount of trips. We have a total of $L_\mathrm{max}=M\times M$ possible links in the system. If we have $N$ trips, which are distributed among individual connections following the probabilities $p_{ij}$, the expected number of links is (see equation 5 in \cite{emigh1983number})
\begin{equation}
    E(L)=L_\mathrm{max}-\sum_{i,j}^M \left\{1-p_{ij}\right\}^N
\end{equation}
which monotonously depends on $N$. Thus, the sampling process accounts for changes in the number of links, while the rescaling approach does not.

In our methodology, we generally use both the resampling approach as well as the rescaling approximation of equation Eq.~\ref{eq:rescaling_wrong} where the approximation is valid. When we correct for data generation, the networks $F'^{(q)}$ can contain both fewer or more trips than the original networks $F^{(q)}$ depending on the ratio $a=\q{N}/N$. As such, we use the resampling approach. For the coverage rates, we find $a=1/c_i \gg 1$ in the vast majority of districts, such that the rescaling approximation is valid.




\section{Comparison of original and debiased networks}
\label{sec:comparison_orig_debiased}

The debiased networks ($\widehat{F}$) differ from the original mobility network ($F$) in many ways.
Fig.~\ref{fig:networks_orig_vs_debiased} show the original and debiased networks for Sierra Leone, DRC, and Iraq, and Table~\ref{tab:debiasing_statistics} includes some key statistics of the networks. 
Most notably from a visible inspection, the debiased networks contain data for all districts in the country, while the original dataset left regions unrepresented. The number of nodes or regions increases in all districts, as does the population that is represented in the networks (for the latter we count the population that is living in the district). Similarly, the number of total trips in the network increases notably in all districts. This is mainly caused by the upscaling of flows with the coverage rate $c_i$ outlined above. The original dataset only represents a small fraction $c_i<1$ of the population, and upscaling the data to the full population increases the numbers considerably. In addition, adding trips from the gravity model increases the total flow in the network, although the effect is smaller, especially in DRC and Iraq where only a smaller number of districts were not present at all in the original data.

The absolute number of links increases for Sierra Leone but decreases for DRC and Iraq. In the debiasing process, the number of links in general decreases due to the resampling process given in Eq.~\ref{eq:resampling_multinomial}, as not all links present in the original network will necessarily be sampled, is unaffected by the rescaling process, and is increased by the imputation of flows from the gravity model, so that overall it can decrease as well as increase.

However, more interesting is the change in the distribution of flow among the links and nodes in the network, see Fig.~\ref{fig:networks_orig_vs_debiased_statistics}. Overall, the distribution of flow $F$ per link is shifted to higher values, as the total number of trips in the network increases due to debiasing. But the distribution is also less wide and has a smaller relative variance, as indicated by the smaller coefficient of variation $c_v$ in the debiased networks. This is true for both the distribution of flows $F_{ij}$ and node strengths $s_{i}=\sum_j F_{ij}$. This change indicates that the number of trips is more evenly distributed among all possible connections as well as the districts.

\begin{figure}[tb]
\centering
\includegraphics[width=0.7\linewidth]{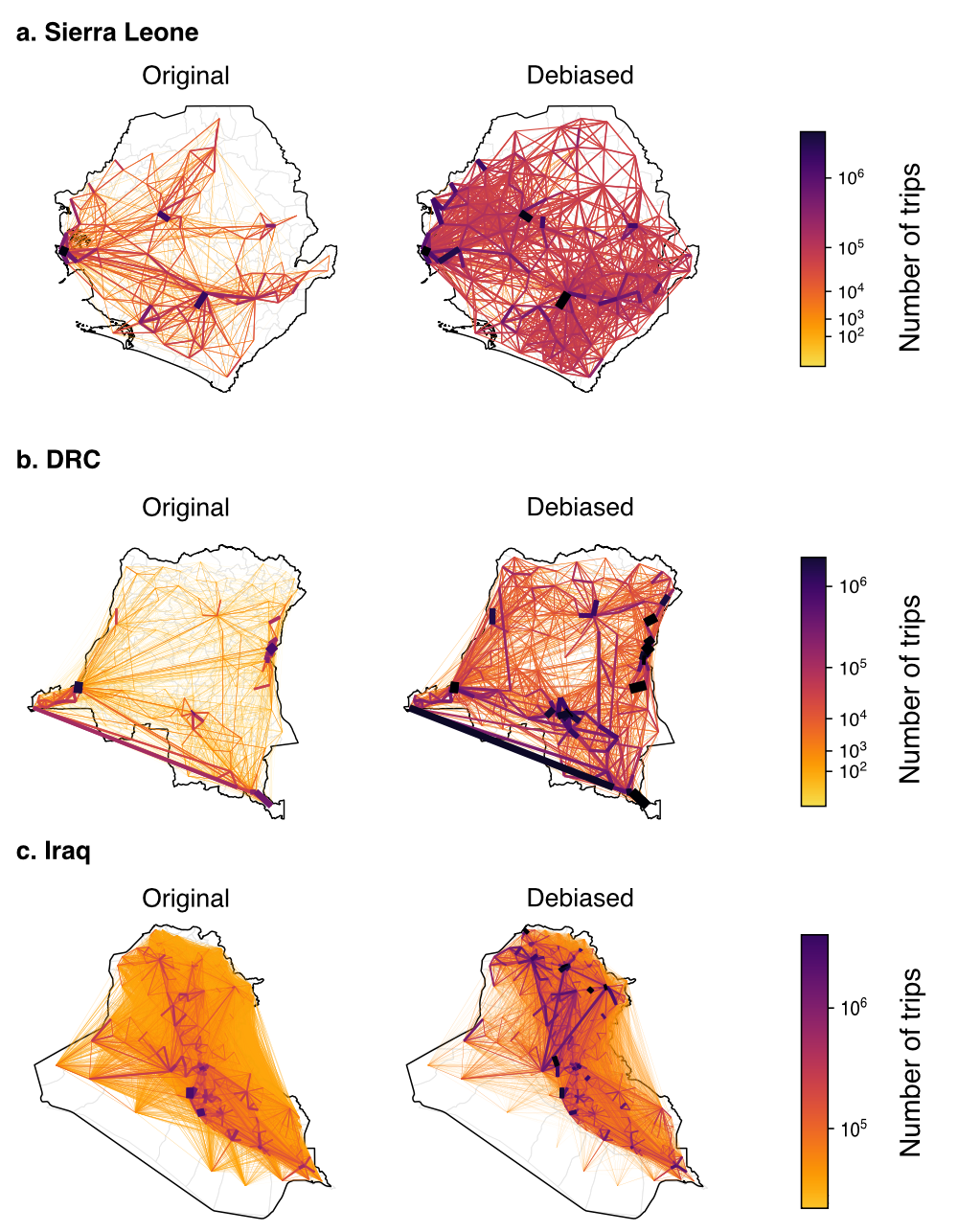}
\caption{\textbf{The original mobility networks compared to the debiased networks.} All networks show marked structural differences due to debiasing. Most visible is the addition of flows from the gravity models, which is less pronounced for the Iraq network as the original network is already very dense. Also visible is the rescaling of flows. In the original network, the flows are centered at cities or densely populated areas, while the network is more decentralized and evenly distributed geographically in the debiased network. The color and width of links is proportional to the flow along each connection, where the scales of both color and widths are tuned for each country separately to make difference more apparent.}
\label{fig:networks_orig_vs_debiased}
\end{figure}

\begin{table}[htb] 
\centering
\begin{tabular}{|c || c | c | c | c | c | c |} 
 \hline
 \rule{0pt}{3ex}
  & \multicolumn{3}{c|}{Nodes/Regions} & \multicolumn{3}{c|}{Links} \\
 \hline
\rule{0pt}{3ex}
 Country & $F$ & $\widehat{F}$ & $\%$ & $F$ & $\widehat{F}$ & $\%$ \\
 \hline
 \rule{0pt}{3ex}
 Sierra Leone & 88 & 153 & +73\% & 1,658 & 2,807 & +69\%\\
 DRC & 132 & 150 & +13\% & 3,304 & 2,700 & -18\% \\
 Iraq & 352 & 393 & +12\% & 90,242 & 75,393 & -17\% \\
\hline
\hline
 \rule{0pt}{3ex}
 & \multicolumn{3}{c|}{Trips} & \multicolumn{3}{c|}{Population included} \\
 \hline
 \rule{0pt}{3ex}
  Country & $F$ & $\widehat{F}$ & $\%$ & $F$ & $\widehat{F}$ & $\%$ \\
  \hline
  \rule{0pt}{3ex}
  Sierra Leone & 1.09 $\cdot 10^9$ & 6.98$\cdot 10^9$ & + 542\% & 4,864,735 & 6,158,361 & + 26.6\% \\
  DRC & 8.74$\cdot 10^8$ & 8.33$\cdot 10^9$ & + 853\%  & 104,156,723 & 108,664,385 & + 4.3\% \\
  Iraq &  1.27$\cdot 10^9$ & 4.60$\cdot 10^9$ & + 263\% & 33,610,757 & 33,970,498 & + 1.07\% \\
 \hline
\end{tabular}
\vspace{1em}
\caption{\textbf{Statistical changes in the network due to debiasing}. We compare the several key statistics of the original mobility network $F$ with a realization of the debiased network $\widehat{F}$. Compared are the number of nodes and links in the network, the total number of trips, and the population in all districts that are represented in the network.}
\label{tab:debiasing_statistics}
\end{table}

\begin{figure}[!htb]
\centering
\includegraphics[width=0.9\linewidth]{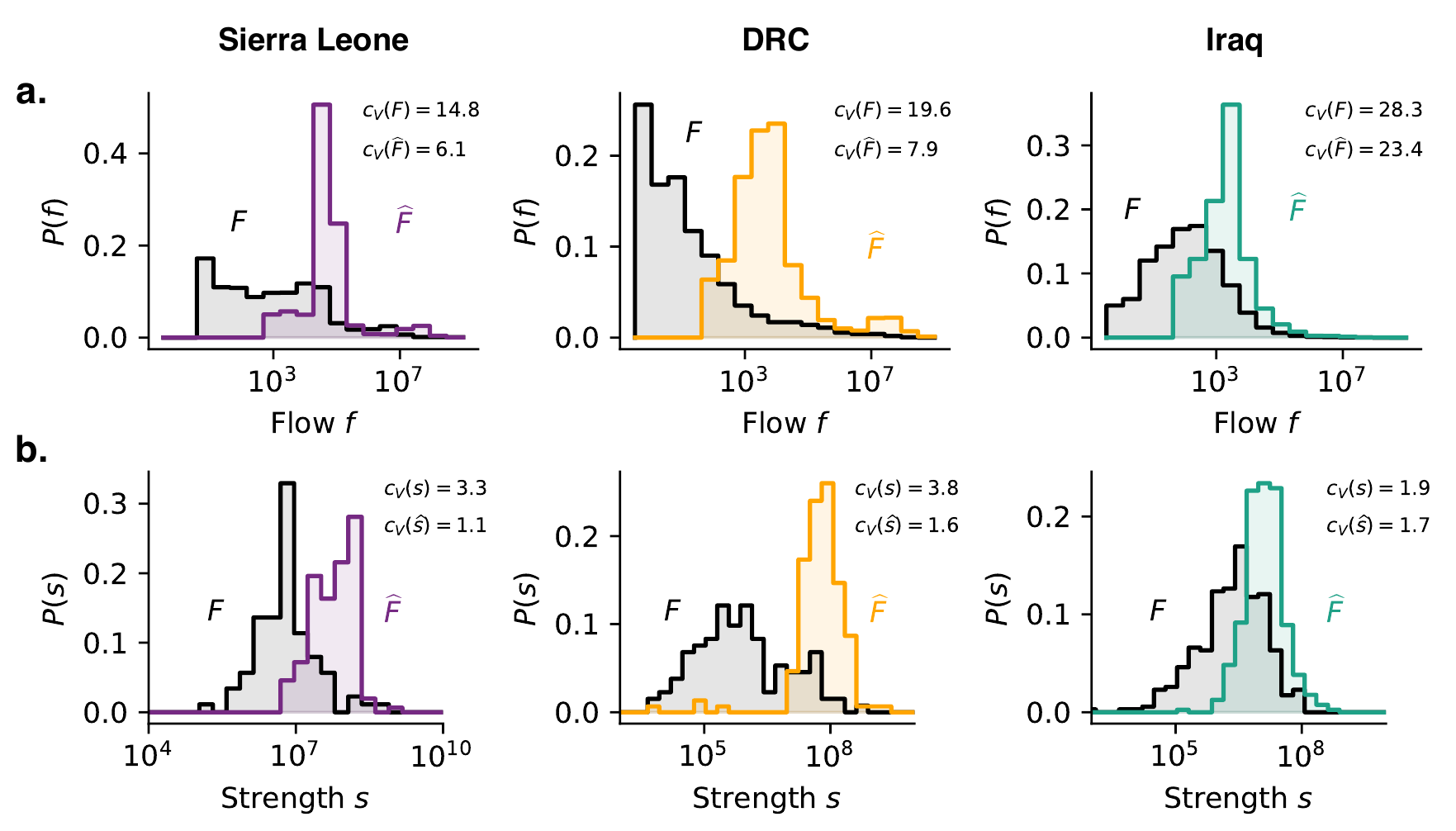}
\caption{\textbf{Differences between the original and debiased networks.} Shown are \textbf{a)} the distribution of flows $f$, that is the number of trips along edges $F_{ij}$, and \textbf{b)} the node strength $s$ in the network, that is the sum of outgoing trips from a district, $s_i=\sum_j F_{ij}$. Each distribution is shown for the original network $F$ (black) and the debiased network $\widehat{F}$ (colored). After debiasing, the distributions are shifted to higher values, as the total flow in the network is increased, especially due to the rescaling of flows with the coverage $c_i$. In addition, the values are distributed more closely together, as shown by the decreased coefficient of variation $c_v(x)=\frac{\sigma(x)}{\overline{x}}$, which indicates that the number of trips is more equally distributed among edges and nodes in the mobility network.}
\label{fig:networks_orig_vs_debiased_statistics}
\end{figure}

\section{SIR model and simulation}
\label{app:SIR}

We implement a metapopulation SIR-model \cite{Ray1997, Keeling2007}, which models epidemic spreading in subpopulations connected by mobility.
The subpopulations correspond to the $M$ districts of the country, where each district $i$ has the population $P_i$ taken from census data. Each individual in the subpopulations belongs to one of the compartments susceptibles $S_i$, infecteds $I_i$, or recovereds $R_i$, such that $P_i=S_i + I_i + R_i$.
Individuals can move between compartments due to mobility. Specifically, we use a type of commuter mobility \cite{Sattenspiel1995, tizzoni2014use} as used in \cite{schlosser2020covid}. A description of the model is also given in the SI of \cite{schlosser2020covid}, which we again provide here for convenience. An implementation of the model in Python is available at \url{https://github.com/franksh/EpiCommute}.

\subsection{Commuter mobility}

The commuter-type mobility assumes that each individual has a home and a work location. Every day, the individual moves from the home to the work location and back (where both locations can also be the same districts). These commuting patterns are captured in the mobility flows $F_{ij}$, as measured from data, which we can use to calculate the number of commuters between each pair of districts. First, we construct the normalized commute probability matrix $p_{ij}$ by normalizing the outgoing flows for each source district $i$,
\begin{equation}
    p_{ij} = \frac{F_{ij}}{\sum_j F_{ij}}
\end{equation}
such that $\sum_j p_{ji}=1$. Further, we define
\begin{equation}
    P_{ij}=p_{ij} \cdot P_i
\end{equation}
as the sub-population of individuals that live in location $i$ and commute to district $j$ for work. Due to the normalization of the probabilities, we have $P_i = \sum_j P_{ij}$.

\subsection{Infection dynamics}

Initially, we set the number of susceptibles $S_i$ in each district equal to the census population $P_i$. In our commuter-mobility framework, the $S_i$ are distributed among the different commuter compartments according to the commuting-probabilities (same as the total population $P_i$),
\begin{equation}
    S_{ij}=p_{ij} \cdot S_i,
\end{equation}
meaning that $S_{ij}$ are those susceptibles that live in $i$ and work in $j$ (the same is true for $I$ and $R$).

We then choose a district $i*$ as the seed of the infection. For the simulations in the main text, we chose the district $i*$ semi-randomly by iterating over all districts and choosing each district in turn. We find that this is decreases the variance of the simulation outcomes. We then set an initial seed of $I_{i^*}(t=0)=100$, which are distributed among the commuter-compartments $ij$ again relative to the probabilities $p_{ij}$ (and we simultaneously decrease $S_{i^*}(t=0)$ by the same amount).

A susceptible in compartment $S_{ij}$ can get infected in two ways:
\begin{enumerate}
    \item while the individual is at home in compartment $i$, by infecteds present in $i$ (including commuters from locations $j$ to location $i$)
    \item while commuting to a compartment $j$, by all the infecteds present in $j$ (including commuters from other locations $k$ to location $j$).
\end{enumerate}
We assume that people spend equal amounts of time at home and commuting. The dynamics of a compartment $S_{ij}$ can then be given as

\begin{equation}
\frac{\mathrm{d}S_{ij}}{\mathrm{d}t} = -S_{ij}\left( \lambda_i^\mathrm{home} + \lambda_j^\mathrm{work} \right) = -S_{ij}\left( 
\frac{\beta}{2}\frac{\sum_k^m I_{ik}}{\sum_k^m P_{ik}} + \frac{\beta}{2}\frac{\sum_k^m I_{kj}}{\sum_k^m P_{kj}}
\right),\label{eq:commuter_S}
\end{equation}
where the first term $\lambda_i^\mathrm{home}$ is the force of infection while at home (where transmission can occur from all the infected at home in $i$), and the second term $\lambda_j^\mathrm{work}$ is the force of infection while commuting (where transmission can occur from all the infected commuting to $j$). 

The dynamics for $I_{ij}$ follow analogously with an additional recovery term, i.e.
\begin{equation}
\frac{\mathrm{d}I_{ij}}{\mathrm{d}t} = - \mu I_{ij} + S_{ij} \left( \lambda_i^\mathrm{home} + \lambda_j^\mathrm{work} \right).\label{eq:commuter_I}
\end{equation}
Since the population size is constant, the third equation follows as $\mathrm dR_{ij}/\mathrm{d}t=\mu I_{ij}$. The disease-specific dynamics are parametrized by the infectivity rate $\beta$ and the recovery rate $\mu$.

\subsection{Stochastic simulation}

Numerical simulations of the model are performed using a stochastic binomial sampling algorithm. From the above equations, it follows that the probability that an individual in compartment $S_{ij}$ becomes infected in the time interval $\left[t, t+\Delta t \right]$ due to the total force of transmission $\lambda_{ij} = \lambda_i^\mathrm{home} + \lambda_j^\mathrm{work}$ is

\begin{equation}
    P(\Delta t; \lambda_{ij}) = 1 - \mathrm{e}^{-\lambda_{ij}\Delta t}.    
\end{equation}

The force of transmission $\lambda_{ij}$ changes over time, but if $\Delta t$ is chosen small enough we can approximate $\lambda_{ij}$ to be constant during $\Delta t$. Consequently, we can determine the number of individuals $S_{ij}$ that become infected during $\left[t, t+\Delta t \right]$ by drawing from a binomial distribution with the probability $P(\Delta t; \lambda_{ij}),$

\begin{equation}\label{eq:S_to_I}
\Delta \left( S_{ij}\rightarrow I_{ij}\right) \sim \mathrm{Binom}(S_{ij}(t), P(\Delta t; \lambda_{ij})).
\end{equation}

Likewise, the amount of infected in $I_{ij}$ that recover during the time is given by

\begin{equation}\label{eq:I_to_R}
\Delta \left( I_{ij}\rightarrow R_{ij}\right) \sim \mathrm{Binom}(I_{ij}(t), P(\Delta t; \mu)).
\end{equation}

For each time step, we compute both Eqs.\ref{eq:S_to_I} and \ref{eq:I_to_R} and then update the system accordingly. The simulation ends when either a maximum time $T_\mathrm{max}$ is reached or the number of infecteds reaches 0.

\subsection{Epidemic simulation results}

In addition to the results of the epidemic results in the main text, Figure.~\ref{fig:SI_epi_results} shows the country-wide epidemic curves for Sierra Leone, DRC and Iraq. Overall we find a less pronounced impact of biases on a country-wide level. For Sierra Leone, the original data underestimates the peak of infecteds as well as the arrival time, while differences are smaller for DRC and Iraq.

\begin{figure}[htb]
\centering
\includegraphics[width=0.9\linewidth]{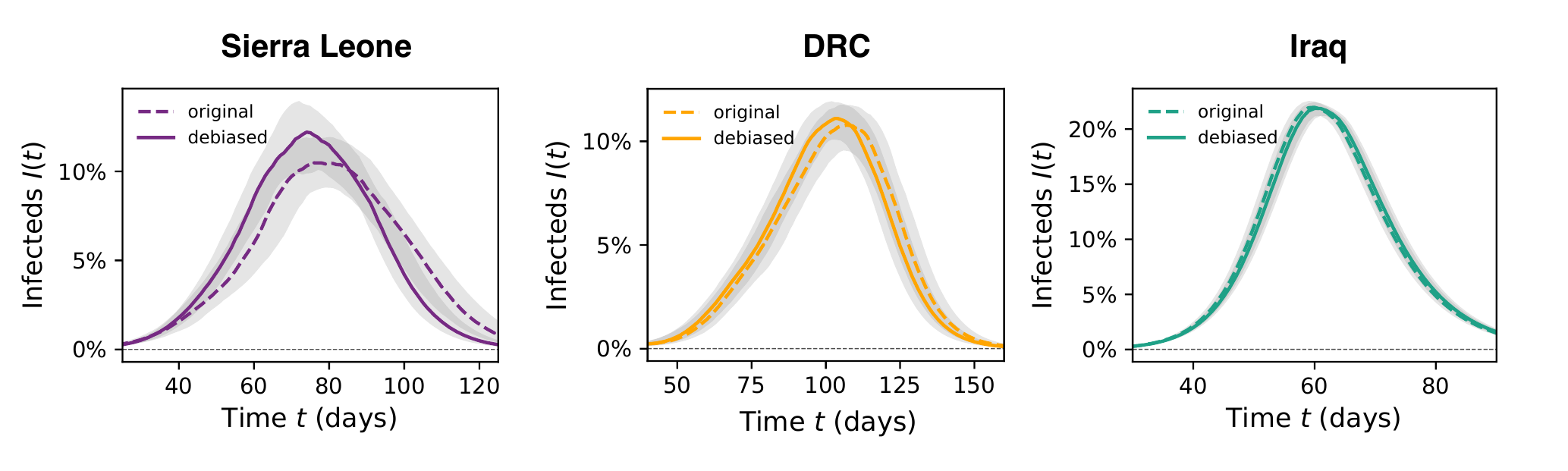}
\caption{\textbf{Additional results of the epidemic simulations. Shown are the epidemic curves of infecteds $i(t)$ on a country-wide level. Lines show the median of $i(t)$, and shaded areas encompass the most central $50\%$ of curves (see Materials and Methods in main manuscript)}}
\label{fig:SI_epi_results}
\end{figure}

\bibliographystyle{unsrt}  
\bibliography{references}